\documentclass[fleqn,usenatbib,useAMS,usedcolumn]{mnras}
\usepackage{graphicx,amsmath}
\usepackage{subfigure}
\usepackage[T1]{fontenc}
\usepackage{ae,aecompl}
\usepackage{txfonts}

\begin{document}

\title[Matching Chemo-Dynamical Simulations to Observations of the Milky Way]
{The Gaia-ESO Survey: Matching Chemo-Dynamical Simulations to Observations of the Milky Way
\thanks{Based on data products from observations made with ESO Telescopes at the
La Silla Paranal Observatory under programme ID 188.B-3002  (The Gaia-ESO Public Spectroscopic Survey).}}

\author[B.~B. Thompson et~al.]{B.~B. Thompson\thanks{email: ben@benjaminbthompson.com},$^{1,2,3,4}$ C.~G. Few,$^{2,4,5}$  M. Bergemann,$^6$ B.~K. Gibson,$^{2,4}$ B.~A. MacFarlane,$^1$
\newauthor A. Serenelli,$^7$ G. Gilmore,$^{8}$ S. Randich,$^9$ A. Vallenari,$^{10}$ E.~J. Alfaro,$^{11}$ T.~Bensby,$^{12}$
\newauthor P. Francois,$^{13}$, A.~J. Korn,$^{14}$ A.~Bayo,$^{15}$ G. Carraro,$^{16}$ A.~R. Casey,$^{8}$ M.~T. Costado,$^{9}$ 
\newauthor P. Donati,$^{17}$ E. Franciosini,$^{16}$ A. Frasca,$^{18}$ A. Hourihane,$^{8}$ P. Jofr\'e,$^{8}$ V. Hill,$^{19}$ U. Heiter,$^{14}$
\newauthor S.~E. Koposov,$^{8}$ A. Lanzafame,$^{18,20}$ C. Lardo,$^{21}$, P. de Laverny,$^{22}$ J. Lewis,$^{8}$  L. Magrini,$^{9}$  
\newauthor G. Marconi,$^{16}$   T. Masseron,$^8$ L. Monaco,$^{23}$ L. Morbidelli,$^{9}$ E. Pancino,$^{9}$ L. Prisinzano,$^{24}$ 
\newauthor  A. Recio-Blanco,$^{22}$ G. Sacco,$^{9}$ S.~G. Sousa,$^{25}$ G. Tautvai\v{s}ien\.{e},$^{26}$ C.~C. Worley,$^{8}$ S. Zaggia,$^{10}$  \\
\\
Affiliations are listed after the references.}

\date{\today}
\pagerange{\pageref{firstpage}--\pageref{lastpage}} \pubyear{2017}
\maketitle
\label{firstpage}

\maketitle

\begin{abstract}
The typical methodology for comparing simulated galaxies with observational
surveys is usually to apply a spatial selection 
to the simulation to mimic the region of interest covered by a comparable observational 
survey sample. In this work we compare this approach with a more sophisticated
post-processing in which the observational uncertainties and selection effects 
(photometric, surface gravity and effective temperature) are taken into account.
We compare a `solar neighbourhood analogue' region in a model Milky Way-like galaxy 
simulated with \textsc{ramses-ch} with fourth release Gaia-ESO survey data.  We find that a 
simple spatial cut alone is insufficient and that observational uncertainties 
must be accounted for in the comparison. This is particularly true when the scale of uncertainty 
is large compared to the dynamic range of the data, e.g. in our comparison, the [Mg/Fe] distribution is 
affected much more than the more accurately determined [Fe/H] distribution. Despite clear differences 
in the underlying distributions of elemental abundances between simulation and observation, incorporating 
scatter to our simulation results to mimic observational uncertainty produces reasonable agreement. 
The quite complete nature of the Gaia-ESO survey means that the selection function 
has minimal impact on the distribution of observed age and metal abundances but this would become 
increasingly more important for surveys with narrower selection functions.
\end{abstract}


\section{Introduction}
\label{intro}

The characteristic abundance ratios found in different stellar populations provide us with an 
opportunity to uncover the history of galaxy formation. Using what is known as 
galactic archaeology to link the chemistry, ages and dynamics of stars allows 
us to trace the origins of the components of the Milky Way \citep{1962ApJ...136..748E, 2002ARA&A..40..487F}.
We have learned a great deal about the processes associated with galaxy formation
using the essential tools of chemical evolution models and simulations of galaxy formation 
 \citep[e.g.][]{2005MNRAS.364..552S, 
 2008MNRAS.385....3S, 2009MNRAS.398..591S, 2011ApJ...729...16K, 2011MNRAS.415..353W, 2012MNRAS.427.1401C, 
2012A&A...540A..56P, 2012A&A...547A..63F, 2013A&A...554A..47G, 2016A&A...587A..10M}
 ands semi-analytic tools \citep{2009MNRAS.400.1347C, 2013MNRAS.435.3500Y}.
More recently, we have gone beyond tracing dynamics and global metallicity within simulations 
to include chemical evolution in such a way that individual elements and isotopes can 
be traced in combination with self-consistent galaxy formation scenarios 
\citep[e.g.][]{1995MNRAS.276..549S, 2002MNRAS.330..821L, 2003MNRAS.340..908K, 2004MNRAS.349L..19T, 2008MNRAS.387..577O, 2014MNRAS.444.3845F}.

Comparison of these chemodynamical models with observed trends is fundamental to establishing the validity 
of the models and understanding the observations. 
 A wealth of high precision observational data is required for such comparisons to be made, and thus detailed testing 
of chemical evolution models is only achievable through comparison to Milky Way stars with external galaxies providing 
only mean trends.
Yet despite improvements to the abundance of observational datasets and in simulation resolution, 
the way in which these comparisons are conducted has remained unaltered for decades.
It is straightforward, and indeed common to simply take the results of a simulation of a
Milky Way-like galaxy and compare it like-for-like with observations of the Milky Way itself.
Typically, a spatial region in a simulation that is similar to the one covered by observational 
data of interest is sampled and the stellar properties are directly compared 
\citep[e.g.][]{2008MNRAS.388...39M, 2011ApJ...729...16K, 2012MNRAS.427.1401C, 2014MNRAS.444.3845F}. 

One strong argument against this simple comparison method
is that it ignores observational 
biases. Firstly, the observed datasets have inherent uncertainties,
either systematic (because of the stellar model atmospheres) or random
(instrumental effects). Secondly, observational surveys usually
observe stars within some range of stellar parameters or distances,
which is usually dictated by the intention to study specific types of stars
(low-mass or high-mass, low- or high-metallicity) in certain Galactic populations. This
selection function \citep{2016MNRAS.460.1131S} creates biases in the
distribution functions of the observed dataset. Most commonly,
selection based on colour and apparent magnitude of stars is reflected
in the shape of the metallicity and age distribution functions \citep{2014A&A...565A..89B}.

In galaxy formation simulations, stellar properties are typically represented by ``star particles'', 
which describe the combined properties of a coeval group of stars (a simple stellar population), 
its total stellar mass and metallicity\footnote{In this work, metallicity is defined as the iron abundance, [Fe/H]}.
 Thus one is limited primarily to the integrated luminosities and averaged chemical composition on the 
scale of open clusters within simulations, i.e one star particle represents the mean properties of an open cluster.

As models improve, the detailed distribution of stellar ages and metallicities - in addition to their mean - become increasingly important. It is thus crucial that the approach to derive ``observables'' from the simulated data for comparison with real observations is as close as possible to the methodology employed by observers.

In this work, we discuss a `solar neighbourhood analogue' region in a model Milky Way-like galaxy simulated with the 
\textsc{ramses-ch} code \citep{2002A&A...385..337T,2012MNRAS.424L..11F, 2014MNRAS.444.3845F}, which is post-processed 
using the \textsc{SynCMD} toolkit \citep{2012A&A...545A..14P} to mimic observational selection functions. 
The simulated data are compared with the Gaia-ESO spectroscopic stellar survey 
\citep{2012Msngr.147...25G, 2013Msngr.154...47R}. The Gaia-ESO survey is the largest ongoing high-resolution 
spectroscopic survey of stars in the Milky Way. In the high-resolution ($R \sim 47\,000$) mode, the goal 
is to acquire spectra for about $5\,000$ field stars, probing distances $\sim$2 kpc from the Sun. Here 
we use the results from the fourth data release of the survey (hereafter, \emph{GES-iDR4}), which includes 
all stellar spectra for the first 18 months of the survey. Our simulated solar neighbourhood analogue 
encapsulates a 2~kpc spherical region of space in our simulated galaxy.

We apply different degrees of post-processing on the simulated data to mimic observational effects. Within 
our simulation data, we sample a spatial region analogous to the solar neighbourhood region covered by 
the Gaia-ESO survey and discuss three different methods of transforming the simulated data into the 
`observer plane'. Our first method is simply the traditional spatial selection of the 
solar neighbourhood analogue. The second method applies a 
scatter to the age, metallicity, and Mg abundance based on the uncertainty in the observed datasets. 
The final method applies \textsc{SynCMD} in addition to the observationally motivated scatter of the 
data to include the survey selection functions. 

The paper is organised as follows. We describe the methodology employed in this work in \S\ref{method} 
and the chemodynamical simulation code used in \S\ref{ramsesch}, the properties of the simulated galaxy 
in \S\ref{selenech} and the \textsc{SynCMD} toolkit in \S\ref{syncmd}. We describe the Gaia-ESO survey data 
used in this work in \S\ref{introges} and the data reduction and the observational survey selection 
function in \S\ref{ges}. We introduce the methods of sampling the simulated solar neighbourhood analogue 
in \S\ref{solar} and expand on that for the unaltered simulated galaxy in \S\ref{sim} convolving 
that with \emph{GES-iDR4} errors in \S\ref{scatter} and the application of \textsc{SynCMD} in \S\ref{syn}. 
Finally we describe our results in \S\ref{results} and conclude with \S\ref{conclusion}.

\section{Methodology}
\label{method}

In this work we use \textsc{SynCMD} to `observe' a cosmologically simulated galaxy 
to demonstrate the significance of different aspects of observational effects 
when comparing models with empirical distributions. In this section we 
describe the simulations, initial conditions and the observations as well as
the post-processing applied to mimic observational effects. 

\subsection{RAMSES-CH}
\label{ramsesch}

Our cosmological simulations are performed with the grid code \textsc{ramses} \citep{2002A&A...385..337T}.
To trace the chemical evolution of the simulated galaxy we employ a chemodynamical 
patch called \textsc{ramses-ch} \citep{2012MNRAS.424L..11F, 2014MNRAS.444.3845F}. \textsc{ramses-ch} 
is able to perform N-body and hydrodynamical simulations including stars, dark matter and gas. 
The adaptive mesh refinement method used in \textsc{ramses} allows for refinement of the grid on a cell-by-cell 
basis increasing the resolution in dense regions of the volume. 
This refinement allows for a reduction in computing time while 
maintaining a high resolution spatial grid around 
the galaxy and also capturing large-scale 
cosmological phenomena. \textsc{ramses-ch} includes treatments of 
self-gravity, hydrodynamics, star formation, supernova feedback, gas cooling and 
chemical enrichment. 

Dense gas cells form star particles if the density surpasses a number 
density threshold of $n_{0}$~=~0.1~cm$^{-3}$. The rate at which
the stellar population particles are produced is  
$\dot\rho_* =\epsilon_*\rho_\mathrm{g}/t_\mathrm{ff}$, where $t_\mathrm{ff}=(3\pi/32G\rho_\mathrm{g})^{1/2}$ 
is the local gas free-fall time, $\rho_\mathrm{g}$ is the gas mass density and star formation 
efficiency, $\epsilon_*$~=~0.01. Our choices of both $n_{0}$ and $\epsilon_*$ are the same as in 
\citet{2014MNRAS.444.3845F} where $\epsilon_*$ is chosen to reproduce the
Schmidt-Kennicutt relation \citep{1959ApJ...129..243S,1998ApJ...498..541K}. The particles used to 
trace the stellar mass phase are commonly and 
colloquially referred to as ``star particles'' by the community, however, they do not represent single stars 
but coeval stellar populations. In the simulation presented here they have a birth mass of 3.3~$\times$~10$^4$~M$_\odot$.
To avoid confusion, we define these coeval stellar populations
  in the simulations as``stellar population particles''. In contrast, the synthetic particles representing
  individual stars (see \S\ref{syncmd}) are defined as ``synthetic star particles''.

Radiative gas cooling in the simulation is metallicity- and density- dependent. 
Cooling rates are calculated assuming photoionisation equilibrium with a redshift 
dependent uniform UV background \citep{1996ApJ...461...20H}. Cooling rates due 
to the presence of metals are calculated from the total metallicity of the gas which 
is interpolated from the \textsc{cloudy} \citep{1998PASP..110..761F} cooling rates at zero and 
solar metallicity at temperatures exceeding 10$^4$~K; colder gas takes its 
metal cooling rates from \citet{1995ApJ...440..634R}. We also employ the delayed 
cooling feedback mechanism from \citet{2013MNRAS.429.3068T} to account for 
the unresolved multiphase nature of the gas and avoid the spurious loss 
of thermal energy following SN feedback. In addition to these prescriptions 
for gas heating/cooling we also impose a polytropic equation of state as a 
temperature floor. 
This temperature floor prevents the gas
  from reaching the low temperatures at which the Jeans' length of
  the gas is unresolved and unphysical fragmentation may occur. The gas is therefore prevented from falling below 
$T_\mathrm{min}$~=~$T_\mathrm{th}(n_g/n_0)^{\gamma-1}$ where $\gamma=2$, $T_\mathrm{th}$~=~188~K, 
and $n_0$~=~0.1~cm$^{-3}$.

\textsc{ramses-ch} allows us to track the elements H, He, C, N, O, Ne, Mg, Si, and Fe from 
their dominant production sites (SNII, SNIa and AGB stellar winds) into the ISM 
where they are advected with the gas flow and become imprinted on the stellar 
population particles. The details of \textsc{ramses-ch} are described fully in 
\citet{2014MNRAS.444.3845F} but we briefly summarise the main components here.
Energetic feedback from both type-Ia and type-II supernovae (SNIa and SNII respectively) 
is included with each SN injecting 10$^{51}$~erg as thermal energy into the local grid cell, 
AGB stars eject their mass passively into the enclosing grid cell.
We use the model B SNII yields of \citet{1995ApJS..101..181W} with a correction applied to the 
yields after \citet{1995ApJS...98..617T} which halves the quantity of Fe produced by massive stars and AGB yields 
are taken from \citet{1997A&AS..123..305V}. We consider stars in the mass range 0.1--8~M$_\odot$ 
to evolve along the AGB while stars with masses 8--100~M$_\odot$ eject mass and energy as SNII.

The mass distribution of stars in each stellar population
particle is determined by the Initial Mass Function (IMF). In this
work we use the  \citet{1955ApJ...121..161S} IMF, where we treat the IMF as a single power law of
 slope -1.35 with lower and upper mass limits of 0.1 and 100 M$_{\odot}$, respectively. 
The number of SNIa per unit initial stellar mass is also determined by the IMF via the number 
of stars with masses 3--8~M$_\odot$ in binary systems with either a red giant or main sequence 
star. The lifetime of these systems is taken as the main sequence lifetime of the 
secondary star \citep{1997A&A...320...41K}. This combination of chemical evolution model parameters 
is described in \citet{2014MNRAS.444.3845F} as model \emph{S55-uM100-IaK} where the impact of the 
choice of IMF and SNIa model on simulations similar to that presented here is also discussed. 
In brief, the particular IMF that is chosen in the model offsets the [O/Fe] ratio by up to 
0.25~dex as a result of the greater number of SNII progenitors created by a top-heavy IMF. The Salpeter IMF 
chosen for this work results in an [O/Fe] and [Fe/H] distribution that lies between the 
two rather extreme IMFs from \citet{1993MNRAS.262..545K} and \citet{2001MNRAS.322..231K}. The \citet{2003ApJ...586L.133C} IMF is 
now occasionally favoured for chemical evolution and galaxy modelling. The similarity of the 
\citet{2003ApJ...586L.133C} IMF to that of \citet{2001MNRAS.322..231K} which is used in \citet{2014MNRAS.444.3845F} 
means that we can estimate that the mean [Mg/Fe] is lessened by only around 0.05~dex through our choice of IMF than 
if we opted for \citet{2003ApJ...586L.133C}.

There are now numerous hydrodynamical simulation codes to choose from: grid- and particle-based 
codes and the recently emerging moving-mesh and meshless approaches \citep{2010MNRAS.401..791S,2015MNRAS.450...53H}.
The strengths and weaknesses of these codes are explored in idealized test cases \citep{2007MNRAS.380..963A, 2008MNRAS.390.1267T, 2008JCoPh.22710040P}, but also in cosmological galaxy formation \citep{2012MNRAS.423.1726S,2014ApJS..210...14K}  and in isolated galaxy discs \citep{2015MNRAS.450...53H, 2016MNRAS.460.4382F}. 
Specifically, a key property of our code, \textsc{ramses-ch} is its ability to capture metal mixing. This is extremely 
pertinent to this work as it directly affects the dispersion in the abundance ratios of 
the gas which becomes imprinted on the stars. In general, grid-based codes
handle metal diffusion better than particle-based codes 
\citep{2012A&A...540A..56P, 2016A&A...588A..21R} which makes \textsc{ramses-ch} a reasonable
choice for the study of the chemical evolution of galaxies.

\subsection{Galaxy Initial Conditions: Selene-CH}
\label{selenech}

We employ a cosmological `zoom-in' simulation technique using \textsc{ramses-ch} to simulate the 
galaxy: `\emph{Selene-CH}'. This galaxy exists in a box 20~$h^{-1}$~Mpc 
in size created with cosmological parameters ($H_0$,~$\Omega_{m}$,~$\Omega_{\Lambda}$,~$\Omega_{b}$,~$\sigma_{8}$)~=~(70~km~s$^{-1}$,~0.28,~0.72,~0.045,~0.8) 
based on the WMAP 7-year results \citep{2011ApJS..192...18K} and the simulation is run to  $z = 0$. The 
adaptive grid can refine up to 17 levels corresponding to a maximum resolution of 218~pc with 
a dark matter particle mass resolution of 5.64~$\times$~10$^{6}$~M$_{\odot}$ and a stellar population 
particle birth mass of 3.3~$\times$~10$^{4}$~M$_{\odot}$.

The galaxy presented here is a chemodynamical resimulation of the \emph{Selene} initial conditions first 
presented in \citet{2012A&A...547A..63F}. The feedback scheme used in simulating \emph{Selene-CH} is 
different to the original version and so, while the galaxy has roughly the same environment and assembly 
history, some differences are to be expected. The evolution of the stellar distribution in this resimulation 
of \emph{Selene} is analysed in \cite{2017MNRAS.464.3580D}.

The galaxy inhabits a dark matter halo with a mass of 5.245$\times$10$^{11}$~M$_{\odot}$ and 
has a stellar mass of $5.603\times 10^{10}$~M$_{\odot}$. The dark matter halo mass may not 
be a precise match to the halo mass of the Milky Way but in a previous study using the initial 
conditions for this galaxy \citep{2012A&A...547A..63F} it is found that assembly history is far 
more significant than small changes in halo mass. \emph{Selene} as a galaxy is selected because of 
its environment (more than 3~Mpc distant from any other haloes more massive than 3$\times$10$^{11}$~M$_{\odot}$) 
and quiescent assembly history, (no major mergers after redshift $z = 1.0$) that mean it lends itself to comparison with the 
Milky Way. The halo and its properties are identified using the AMIGA halo finder \citep{2009ApJS..182..608K, 2004MNRAS.351..399G}. 
The assembly history of the original version of \emph{Selene} is described in \citet{2012A&A...547A..63F} and more extensively with 
relation to the effect of its assembly on the metallicity and age distribution in \citet{2016A&A...586A.112R}.

\begin{figure*}
\centering
\includegraphics[width=0.75\textwidth]{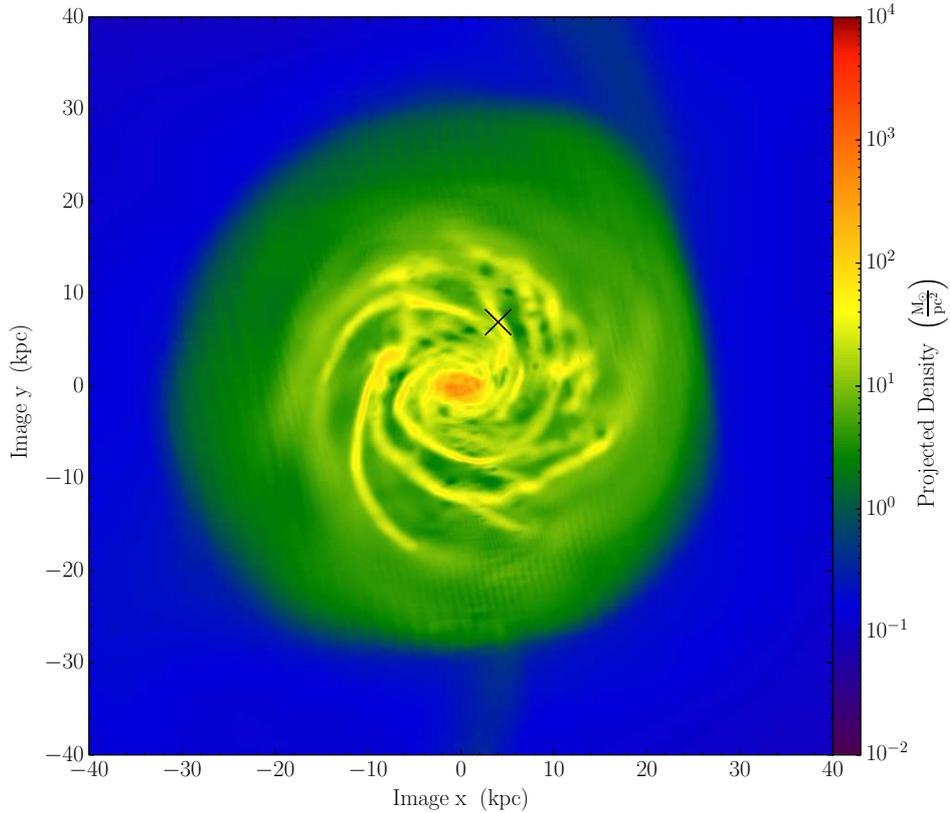}
\caption{Face on gas surface density projection of \emph{Selene-CH}. The galaxy was visualised using the 
\textsc{YT} visualisation toolkit \citep{2011ApJS..192....9T} with a projection depth of 20 kpc. 
The black cross at $x$~=~$4.0$ kpc and $y$~=~$6.93$~kpc is the position of our simulated observer 
as described in Section \S\ref{sim}. The region selected in this work is a 2~kpc sphere around 
the indicated position, similar to the coverage of the Gaia-ESO survey.}
\label{FIG.3}
\end{figure*}

A gas surface density projection of \emph{Selene-CH} is shown in Figure \ref{FIG.3} demonstrating the 
presence and shape of the spiral arms. The cross at $x$~=~4.0~kpc and $y$~=~6.93~kpc is the region where
we place our simulated observer as described in Section~\ref{sim}, 8~kpc from the galactic centre in 
a spiral arm. We use stars from a spherical region 2~kpc in radius around this point which is treated as 
our simulated solar neighbourhood analogue. The size of this region is discussed in Section~\ref{sim}.

\subsection{\textsc{SynCMD}}
\label{syncmd}

The \textsc{SynCMD} synthetic stellar populations generation tool \citep{2012A&A...545A..14P} is a toolkit
 designed to examine simulation data in a similar manner to how an observational survey would sample 
real life stellar populations. The toolkit is used to apply observationally-motivated selection functions to 
simulated stellar population particles. As discussed in \S\ref{ramsesch}, each such particle represents 
a coeval mono-abundance stellar population, its mass is simply the total stellar mass.
The details of the \textsc{SynCMD} code are given in \cite{2012A&A...545A..14P},
and we summarise the process here. A preliminary
application of SynCMD has been undertaken using a RAVE-like selection function \citep{2014nic..confE.149M}.

\textsc{SynCMD} allows us to split a stellar population particle into individual stars by stochastically
 populating a colour-magnitude diagram (CMD). Due to resolution limits in simulations, 
stellar population particles represent `averaged' stellar 
populations rather than individual stars. Stellar population particles typically have a mass of 
 $\sim$10$^4$--10$^6$~M$_{\odot}$ and therefore a stochastic approach is valid. 
We consider each stellar population particle to consist of $10^5$ synthetic star particles, which 
sample the IMF from a mass of 0.15 M$_{\odot}$ to the main sequence turnoff mass at the current age of 
the stellar population particle, up to a maximum of 20 M$_\odot$. All stellar population particles consist of this number of synthetic 
star particles and we weight their contribution to distribution functions by the initial mass of the stellar population 
particle and by the mass remaining in that particle as a fraction of the initial mass, calculated by integrating over
the \cite{1955ApJ...121..161S} IMF.

The physical properties of the synthetic stellar particles are based on theoretical 
stellar models. These properties are $T_{\rm eff}$, log($g$), magnitudes in different 
colour bands, ages, metal abundances and masses. Here we use 
\citet{2008A&A...484..815B,2009A&A...508..355B} isochrones, which cover a 
wide grid of helium and metal abundances ($Y$ and $Z$ respectively), an enrichment 
ratio $\Delta Y / \Delta Z$, and include mass loss by stellar wind and the thermally 
pulsing AGB phase \citep{2007A&A...469..239M}. 
The isochrones are used to calculate a database of simple stellar populations using 
a modified version of \textsc{yzvar}, which has been used in many studies
 \citep[for instance][and references therein]{2003AJ....125..770B}. 
We place the observer in a region of the simulated galaxy analogous
 to the location of the Sun in the Milky Way and generate synthetic
 stars that trace a synthetic CMD by
linearly interpolating in age and metallicity between isochrones of 
simple stellar populations. The interpolation is described in \citet{2012A&A...545A..14P}.

This methodology enables each individual stellar population particle to be mapped to $10^5$ synthetic star particles. 
The mean stellar properties of age, metallicity and metal abundance are that of the parent stellar population particle.
The synthetic star particles are allocated masses at random from the chosen IMF. The masses of these particles are then
used to populate an isochrone using the metal abundance and age of the original stellar population particle from the simulation data.
Properties of the synthetic star particles are calculated from the database and from the stellar population particle's age, metallicty and distance from
the observer. 
The synthetic star particle's properties that are calculated are
the age, luminosity, $T_{\rm eff}$, log($g$), metal abundance,
H abundance, He abundance and magnitudes in the UBVRIJHK bands.
Photometric colours and
magnitude values for each synthetic star will be adjusted according to 
the distance of the star to the simulated observer. The stars retain the age and chemical 
abundances of the simulation particle they are created from.  We eliminate synthetic star particles from the sample that do not fall within the
selection criteria. For comparison with an observational dataset, one can apply a selection function
to the synthetic star particles of the observational survey that the user wishes to emulate. The remaining stars
are used to analyse whatever physical property one wishes, having essentially removed the fraction of each stellar population particle that
would not lie within the selection functions.

\begin{figure}
\centering
\includegraphics[width=0.49\textwidth]{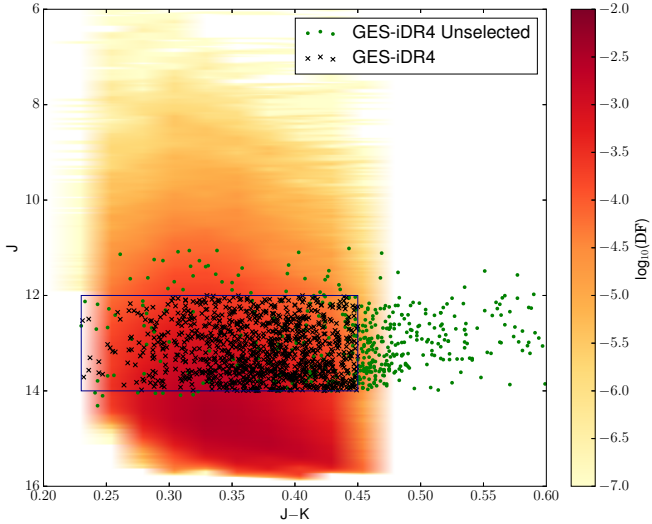}
\caption{Synthetic CMD of J vs J$-$K in apparent magnitude space of the simulated solar neighbourhood analogue
and of stars from \emph{GES-iDR4}. 
The red heatmap represents the synthetic stellar populations from \emph{Selene-SYN} which is 
derived from the simulated galaxy \emph{Selene-CH} as shown in Fig \ref{FIG.3}. The 
black crosses represent stars selected from the \emph{GES-iDR4} dataset 
whereas the stars labelled with green circles representing those removed from 
\emph{GES-iDR4} by the selection function as described in \S\ref{ges}. 
The blue rectangle highlights the J and J$-$K region selection function
 boundary conditions of 12~$<$~J~$<$~14 and 0.23~$<$~J~$-$~K~$<$~0.45.
 Both datasets include the application of surface gravity and effective temperature filters of; 
3.5~$\leq$~log($g$)~$\leq$~4.5~dex and 5400~$\leq$~$T_{\rm eff}$~$\leq$6400~K.
A reddening correction is applied when selecting the observed targets in \emph{GES-iDR4} 
(J$-$K~$+$~$0.5$~E(B$-$V) which we describe in \S\ref{ges}. }
\label{FIG.7}
\end{figure}

\subsection{The GAIA-ESO Survey}
\label{introges}

In this work, we focus on the high-resolution \emph{GES-iDR4} UVES data of the field stars, for which accurate effective temperatures $T_{\rm eff}$, surface gravities log($g$), [Fe$/$H], and Mg abundances are available. These targets were chosen according to their colours to maximise the fraction of un-evolved foreground (FG) stars within 2 kpc in the solar neighbourhood \citep[see][for more details on target selection]{2016MNRAS.460.1131S}. The selection box was defined using the 2MASS photometry \citep{2006AJ....131.1163S,2012ApJS..199...26H}: $12$~$<$~J~$<$~$14$ and $0.23$~$<$~J$-$K~$<$~$0.45$~$+$~$0.5$E(B$-$V). Application of the colour excess E(B$-$V) and target selection are described in \citet{2016MNRAS.460.1131S}. According to these selection criteria, the majority of stars are FG stars with magnitudes down to $V = 16.5$.

For the analysis of the spectra, several state-of-the-art spectrum analysis codes are used \citep{2014A&A...570A.122S}. The observed spectra were processed by 13 research groups within the Gaia-ESO survey collaboration with the same model atmospheres and line lists \citep{2015PhyS...90e4010H}, but different analysis methods: full spectrum template matching, line formation 
on-the-fly, and the equivalent width method. The model atmospheres are 1D LTE spherically-symmetric (log($g$)~$\leq$~$3.5$~dex) and plane-parallel (log($g$)~$\geq$~$3.5$~dex) MARCS \citep{2008A&A...486..951G}. The final parameter 
homogenisation involves a multi-stage process, in which both internal and systematic errors of different datasets are carefully investigated. Various consistency tests, including the analysis of stellar clusters, benchmark stars with interferometric and asteroseismic data, have been used to assess each group's performance. A comprehensive description of the pipelines and tests of the UVES results can be found in \citet{2014A&A...570A.122S}, specifically sections 4, 5, 6, and 7 of that paper.

\begin{figure}
\centering
\includegraphics[width=0.5\textwidth]{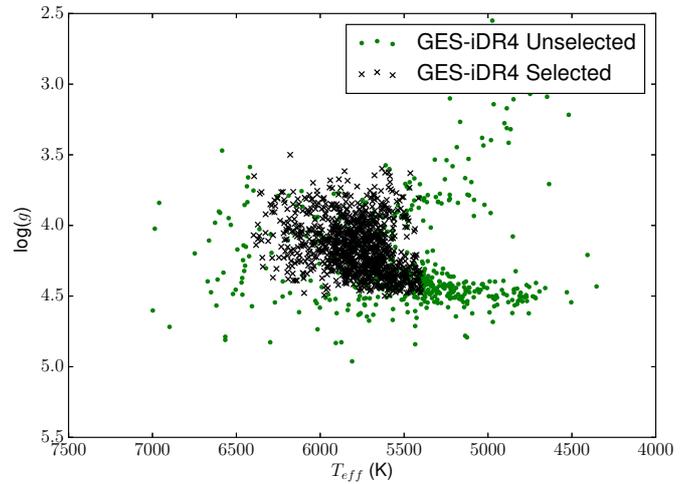}

\caption{$T_{\rm eff}$ vs $\log(g)$ for the \emph{GES-iDR4} observed sample plotted as black crosses. For reference, we also include the stars 
which are removed \emph{GES-iDR4} dataset due to falling outside the selection criteria as green circles. The selection function is described in 
section~\S\ref{ges}.}
\label{FIG.4}
\end{figure}

The final stellar parameters are median of the multiple determinations, and the uncertainties of stellar parameters are median absolute deviations, 
which reflect the method-to-method dispersion. For most stars, the uncertainties are within $100$~K in $T_{\rm eff}$,  $0.15$~dex in log($g$), 
and $0.1$~dex in [Fe/H] and Mg abundances. This accuracy could be achieved because of very careful selection of diagnostics features, very broad 
wavelength coverage and good signal-to-noise ratio of the observed spectra, and validation of the results on the accurate stellar parameters and 
NLTE estimates of chemical abundances of the Gaia Benchmark stars \citep{2015A&A...582A..81J}. In this work, we focus on the high-resolution 
UVES data of the field stars. UVES is the ultraviolet and visual cross-dispersed echelle spectrograph installed at the second unit telescope of 
the VLT \citep{2000SPIE.4008..534D}.  The stars were observed using the UVES U-580 setting, which covers the wavelength 
range from 480 to 680~nm, with a small beam-splitter gap at 590~nm. Most spectra have signal-to-noise ratio between 30 and 100 per pixel. For these 
stars, accurate effective temperatures $T_{\rm eff}$, surface gravities log($g$), [Fe/H], and Mg abundances are available.
The distribution of the \emph{GES-iDR4} sample in the colour-magnitude plane is shown in  Figure~\ref{FIG.7}. 
The $T_{\rm eff}$-log($g$) diagram of the \emph{GES-iDR4} sample is shown in Figure~\ref{FIG.4}.

The ages and masses were determined using the Bayesian code BeSPP \citep{2013MNRAS.429.3645S}. We use the grid of input stellar evolution models computed using the GARSTEC code \citep{2008Ap&SS.316...99W}; it covers a wide range of masses, 0.6~$\leq$~M$_{\odot}$~$\leq$~1.4~M$_{\odot}$ in steps of 0.01~M$_{\odot}$, and metallicities, -5~$\leq$~[Fe/H]~$\leq$ +0.5~dex. The models more metal-poor than -0.6 dex assume  $\alpha$-enhancement of 0.4 dex. Distances were computed using the 2MASS photometry. The ages of the stars are computed from the mode of the posterior PDF.  The uncertainties in age were determined as $\pm$34$\%$ around the median value.

\subsection{Survey Selection Function}
\label{ges}

Before applying the observational selection function to the galaxy simulation, we further post-process the \emph{GES-iDR4} data for the analyses 
presented here. We require that the observed stars must satisfy the following selection criteria for them to be included in the dataset:

\begin{enumerate}
\item The star is not a member of a star cluster and belongs to the Milky Way field population (using the tag 'GES\_MW' in the \emph{GES-iDR4} catalogue);
\item The star is not tagged as a member of a special field such as the CoRoT asteroseismic targets or deep fields in the Galactic centre and anti-centre directions;
\item J-band magnitude $12.0 \leq J \leq 14.0$;
\item J$-$K colour of $0.23$~$\leq$~J$-$K~$\leq$~$0.45$ $+$~$0.5$~E(B$-$V);
\item Heliocentric radial distance of r $\leq 2.0$ kpc;
\item Surface gravity of $3.5 \leq \log(g) \leq 4.5$ dex;
\item Effective temperature of $5400 \leq T_{\rm eff} \leq 6400$ K.
\end{enumerate}

The colour-magnitude cuts were used to create the input catalogues for the Gaia-ESO survey to maximise the number of un-evolved FG stars within 2 kpc in the solar neighborhood.

The $T_{\rm eff}$ and the log($g$) fields of the selection function are chosen because ages of stars with $T_{\rm eff} \loa 5400$ may not be accurate. Likewise, stellar ages are not well determined for hotter stars with $T_{\rm eff} \goa 6400$~K or log($g$) $\goa 4.5$~dex, and for more evolved stars on the red giant branch, log($g$)~$\loa 3.5$~dex. Our selection would thus include subgiants and main-sequence dwarfs.
After applying the selection function as described, we have 1024 stars and use this as our definitive \emph{GES-iDR4} dataset.

\subsection{Analysing the Simulated Solar Neighbourhood}
\label{solar}

The aim of this work is to best examine how different ways of processing the same simulation data give variations in the results obtained 
for the distribution of chemical abundances in the solar neighbourhood analogue. A first-order approach that is commonly used is to simply 
take a spatial region within a simulation that matches the region of interest in a galaxy and compare that with observational data
\citep[e.g][]{2001ApJ...554.1044C, 2008MNRAS.388...39M, 2012A&A...547A..63F, 2014MNRAS.444.3845F}. 
This approach samples the entire stellar population and requires volume completeness for each type of stars, something that no observational 
survey does. The fact that simulations are not subject to observational errors is also usually ignored. As a result, the observed and simulated 
distributions are not directly comparable.

In this work, we compare the  GES-iDR4 results, after the survey selection function is applied, with the following variants of the simulated galaxy \emph{Selene-CH} in order to demonstrate the influence of each component of the process used to mimic observational limits;

\begin{itemize}
\item \emph{Selene-CH} is the unaltered and unmodified galaxy. We select all of the stellar population particles that reside within a 2~kpc sphere around the simulated observer, 15562 in total. These particles are compared directly with the selected \emph{GES-iDR4} results. This kind of direct comparison demonstrates the methodology employed in the `traditional sense', i.e with spatial cuts alone.

\item \emph{Selene-GES} is a modified version of \emph{Selene-CH}. In this case we apply stochastic scattering to the ages and abundance ratios of the stellar population particles to emulate observational uncertainties. The magnitude of the scattering is based on the mean errors taken from the GAIA-ESO dataset.

\item \emph{Selene-SYN} is the result of applying the \textsc{SynCMD} toolkit, as described in \S\ref{syncmd}, to the scattered stellar population particles ages and metallicities in \emph{Selene-GES} (i.e the statistically scattered results of \emph{Selene-CH}). This dataset includes the application of selection functions for log($g$), $T_{\rm eff}$, J-band magnitude, and J$-$K colour and is a more rigorous attempt to mimic the \emph{GES-iDR4} data. 
\end{itemize}

In short, \emph{Selene-CH} represents a first-order analysis of the simulations similar to that
found in the majority of the literature, \emph{Selene-GES} shows the effect of applying observational scatter,
and \emph{Selene-SYN} demonstrates the influence of selection effects. We now describe the post-processing used to create each of these datasets in detail.

\subsubsection{Selene-CH: The `Standard' Simulation Approach}
\label{sim}

We identify a solar neighbourhood analogue within the simulated galaxy \emph{Selene-CH}, a position 
8~kpc from the galactic centre on a spiral arm. This position is shown with
a cross in Figure \ref{FIG.3} at ($x$,$y$) = (4.0, 6.93)~kpc; the solar neighbourhood analogue
is centred on this point relative to the galactic centre. We have repeated the analysis that follows with stars from 
different positions on a circle with a galactocentric radius of 8~kpc and find that our results are 
robust to changes in the position of the simulated observer. This is due to azimuthal 
homogeneity in the age and chemical abundances; the mean azimuthal variations at 8~kpc
from the galactic centre for [Fe/H] and [Mg/Fe] are 0.02~dex and 0.005~dex respectively 
and the mean age variation is only 0.5~Gyr. This local robustness to changes in position means 
that attempts to imitate the right ascension and declination distribution of the GES-iDR4 data would only 
reduce the number of stellar population particles selected and increase noise in the simulated sample.

Previous works have applied different variations of \emph{a posteriori} re-normalisations 
\citep[e.g][]{1995MNRAS.276..505P,2004A&A...421..613F,2010ApJ...724..748H} and/or employed GCE models to 
infer revised sets of stellar yields \citep{2004A&A...421..613F}. We have followed 
this method with our element abundances normalised to the \citet{2009ARA&A..47..481A} solar values and then
further shifted by $\Delta$[Fe/H]~=~-0.066~dex and $\Delta$[Mg/Fe]~=~0.078~dex. This shift in the 
abundance ratios brings the average abundance of the stars aged between 4.0 and 5.0 Gyr to the solar abundance.
This may seem arbitrary, however the amount by which we normalise is small compared to the width 
of the distribution and we are primarily concerned with the dispersion of the element ratios.
Furthermore, it is demonstrated in \citet{2014MNRAS.444.3845F} that variations in abundance ratios 
(particularly those of $\alpha$-elements to Fe) are effectively shifted in the same way depending on the 
IMF. The need to apply such a shift implies that the sub-grid chemical evolution model is not quite 
correct which is hardly surprising given the uncertainties in the underlying yields and chemical 
evolution model. Therefore, while the renormalisation of abundance ratios introduces a slight inconsistency to the 
model it by no means negates our results. 

\subsubsection{Selene-GES}
\label{scatter}

This dataset extends the methodology described above to generate \emph{Selene-CH} by applying a stochastic scattering based on the \emph{GES-iDR4} 
error bars for age, metallicity, and [Mg/Fe] abundance ratio, to mimic the effect of the unavoidable uncertainties found in observations 
on the precisely known (but not necessary accurate) simulated values.

We degrade the precision of our simulated metallicity and [Mg/Fe] data on a particle-by-particle basis using a Gaussian distribution, 
centred on the original simulated value with a standard deviation equal to the mean error found in the \emph{GES-iDR4} dataset: 
$\sigma_{\mathrm{[Fe/H]}}$~=~0.101~dex and $\sigma_{\mathrm{[Mg/Fe]}}$~=~0.120~dex. New abundance ratios for each stellar population 
particle are chosen randomly from this distribution.

The age value for each stellar population particle is also scattered this way except that the distribution from which the new value is 
chosen at random is not symmetric. The ages of the observed stars in the \emph{GES-iDR4} dataset have a mean lower age error of 
$\sigma_{age,l}$~=~3.20~Gyr and a mean upper age error of $\sigma_{age,u}$~=~2.37~Gyr. We construct a piecewise function from 
two half-Gaussians with these standard deviations respectively to scatter the simulated ages. This process not only broadens 
the distributions but also {makes stellar population particles slightly older.

\subsubsection{Selene-SYN}
\label{syn}

Our final version of \emph{Selene} takes the scattered stellar population particles from \emph{Selene-GES}
 and inputs those particles to \textsc{SynCMD} creating a third dataset referred to here as \emph{Selene-SYN}. 
The mechanics of \textsc{SynCMD} are described in \S\ref{syncmd} but the key here is to split the stellar
 population particles into individual synthetic star particles with a realistic distribution of star properties so
 that we can apply photometric, log($g$) and $T_{\rm eff}$ cuts to exactly mimic the observed \emph{GES-iDR4} dataset.
 The selection criteria are stated in \S\ref{ges} however we do not apply the dust extinction correction
 to the J$-$K upper limit because our synthetic star particles are unaffected by dust. The synthetic star particles that 
remain after this are used as our sample of stars analogous to the \emph{GES-iDR4} dataset so that we can 
compare the simulations in a more like-for-like manner. The CMD for \emph{Selene-SYN} is shown in figure \ref{FIG.7}.

As stellar population particles represent different masses of stars, the contribution of each one 
in terms of synthetic star particles is weighted by the initial mass of the stellar population particle
to correctly account for the mass. The initial mass is used because any stars that have evolved and no
longer form part of the stellar population are removed from the $10^5$ synthetic star particles. The 
distribution functions shown in this work are described as `mass-weighted', this means that we have weighted 
the \emph{Selene-CH} and \emph{Selene-GES} particles by their mass to be consistent with the \emph{Selene-SYN}
distribution function.

\section{Results and Discussion}
\label{results}

We now present and discuss the impact that `observing' our simulations has on 
the distribution of selected stars in age, [Fe/H], and [Mg/Fe] in comparison with \emph{GES-iDR4} data.

\begin{figure}
\centering
\includegraphics[width=0.49\textwidth]{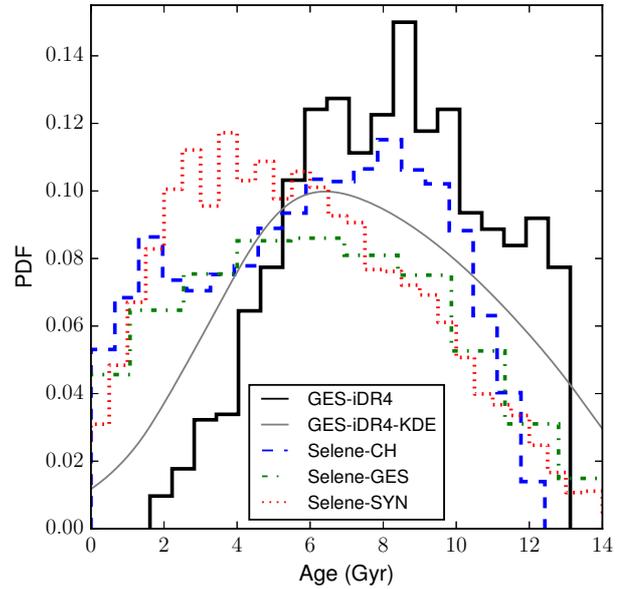}
\caption{A plot of the present day (z=0) age distribution functions for the datasets discussed. The
Gaia-ESO DR4 data (\emph{GES-iDR4}) is shown as a black line, the grey line is a kernel density estimate of the 
\emph{GES-iDR4} dataset to represent the uncertainties on each point; this has the net effect of shifting the distribution towards
younger stellar ages. The simulation datasets \emph{Selene-CH}, 
\emph{Selene-GES} and \emph{Selene-SYN} are shown as blue dashed, green dot-dashed and red triple-dot-dashed lines respectively.}
\label{agedf}
\end{figure}

\subsection{Ages}
We begin the discussion with the analysis of the age distribution in the observed and simulated datasets, 
however, we remind the reader that age determinations for the observed stars are notoriously difficult,
 because they rely on the knowledge of surface stellar parameters, metallicities, and $\alpha$-element abundances 
(Section \S\ref{introges}). Typically, ages of stars in \emph{GES-iDR4} have an uncertainty of  $\sim$$30\%$, which is a statistical error and 
does not include any systematic component. Systematic errors cannot be easily quantified, because of the 
complex interdependence of different parameters and correlated errors (for example, the error in [Fe$/$H] is 
correlated with the error in $T_{\rm eff}$ and in log$(g)$). Therefore, some mismatch between the observed and 
model datasets is expected and should not be taken as the evidence of the failure of the galaxy simulations.

The age distributions of our three versions of \emph{Selene} and the \emph{GES-iDR4} stars are shown in Figure~\ref{agedf}. 
Clearly, there is a systematic difference between the \emph{GES-iDR4} and the simulation data, with an obvious offset to younger stars seen in the simulated data.
The application of the stellar age scattering to the simulated data (\emph{Selene-GES}) has the effect of 
flattening the somewhat truncated older part of the age distribution, removing the peak at 8--10~Gyr and reducing the number of young stars which is entirely expected from the sharp edge of the underlying distribution. Finally, when sampling the CMD of the stellar population particles and applying the \emph{GES-iDR4} selection criteria to the scattered data (\emph{Selene-SYN}) we find that the old end of the age distribution is unaffected, but that the GES photometric filters have the effect of producing a peak between 2--5~Gyr and removing many of the stars with ages below 1~Gyr from the distribution. The latter effect brings the young end of the distribution function closer to the observed distribution but there is still a significant discrepancy in both the shape and mean age. To show how the uncertainties associated with the \emph{GES-iDR4} observations alter the shape of the distribution we have used kernel density estimation to calculate the 
uncertainty weighted probability distribution function, shown as \emph{GES-iDR4-KDE} in Figure~\ref{agedf}. This distribution is not the correct one with which to compare the processed 
simulations and we provide it simply for accuracy in representing the observed data.

The differences in the age distributions could be caused by several effects. Firstly, this could be due to the differences in the underlying distributions of stellar parameters in the observed and simulated samples. In particular, the combination of the SFH from \emph{Selene-CH} and the SSPs database used in \textsc{SynCMD} produce a temperature distribution with a $\sim$400~K hotter mean $T_{\rm eff}$ value than our chosen \emph{GES-iDR4} sample. This is a very significant difference and is most likely central to understanding of the discrepancy, but currently, we have no suitable framework to explore the effect. If the spectroscopic determinations of T$_{\rm eff}$ are biased, this could explain the deficiency in age for young stars in the \emph{GES-iDR4} sample. \citet[][Figure 2]{2014A&A...565A..89B} showed that T$_{\rm eff}$ measurements, especially for stars with T$_{\rm eff} >$ 6000~K, appear to be over-estimated when compared to the more accurate methodology (infra-red flux method). If this also holds true for \emph{GES-iDR4} datasets then by imposing a T$_{\rm eff}$ cut of 6400~K we actually remove stars, which may have even lower T$_{\rm eff}$ and this pushes the observed distribution towards colder (and older) stars.

One should keep in mind that typically, the mean age of stars is a function of galactocentric radius and \emph{Selene-CH} 
does form inside-out \citep[see][]{2012A&A...540A..56P}. This means that the age distribution of stars would shift to 
older values with decreasing radii, and thus a more appropriate 
solar neighbourhood analogue may exist for this galaxy, however given our uncertainty regarding the true 
distribution we have opted to select our region of interest based on the distance from the galactic center. 
Finally, the discrepancy in the age distributions could be due to the differences in the star formation history. 
Currently, we have no robust constraints on the star formation history of the Milky Way disc over the past 10 Gyr 
and a detailed analysis of this very complex problem is beyond the scope of this work.

\begin{figure*}
\centering
\includegraphics[width=1.0\textwidth]{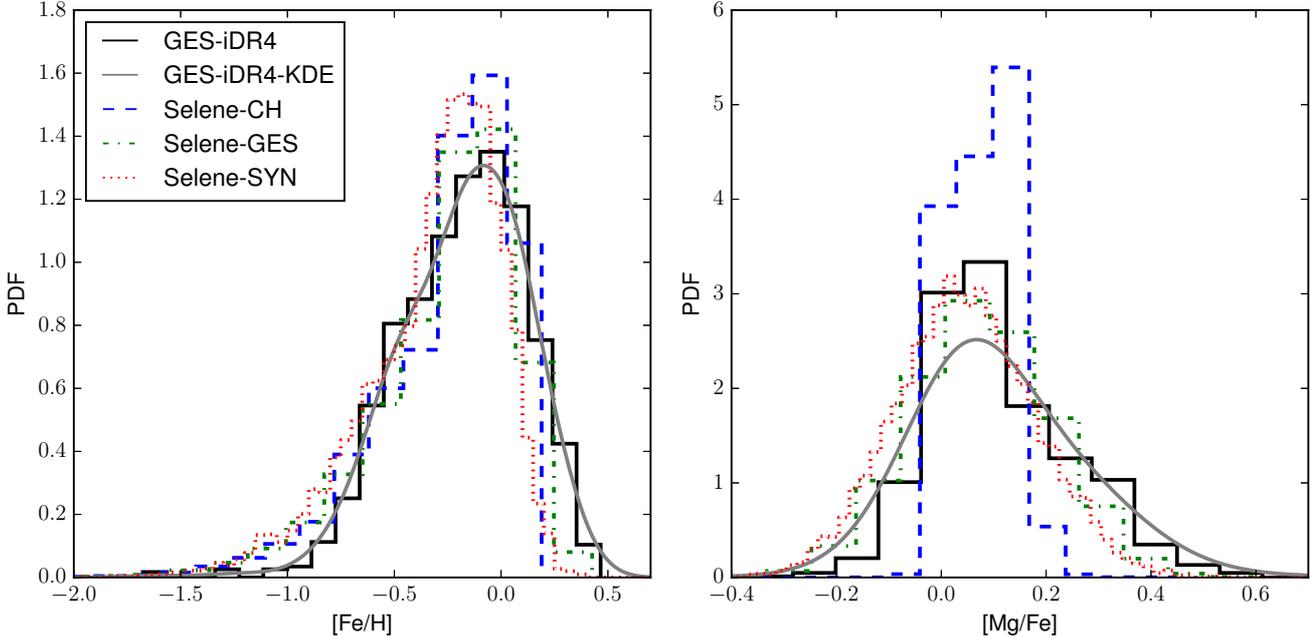}
\caption{[Fe/H] and [Mg/Fe] distribution functions in the solar neighbourhood analogues of the mass-weighted \emph{Selene-CH} 
and \emph{Selene-GES} datasets, the synthetic observation dataset \emph{Selene-SYN} and the observation dataset \emph{GES-iDR4}. 
\emph{Selene-CH}, \emph{Selene-GES} and \emph{Selene-SYN} represented as blue dashed, green dot-dashed and red triple-dot-dashed lines respectively. 
The distribution functions of the \emph{GES-iDR4} stars are shown in black, the grey line is a kernel density estimate of the
\emph{GES-iDR4} dataset to represent the uncertainties on each point; this has little effect on the [Fe/H] distribution but does diminish the 
peak in [Mg/Fe]. The bin widths for \emph{Selene-CH}, \emph{Selene-GES} and \emph{Selene-SYN} 
in [Fe/H] is 0.05~dex, and for \emph{GES-iDR4} this is 0.1 dex. The bin widths for \emph{Selene-CH} and \emph{GES-iDR4}  
in [Mg/Fe] is 0.05 dex, for \emph{Selene-GES} and \emph{Selene-SYN} is 0.02 dex.}
\label{mdf}
\end{figure*}

\subsection{Distribution functions of [Fe/H] and [Mg/Fe]}
Figure~\ref{mdf} shows the distribution functions of [Fe/H] (left-hand panel) and [Mg/Fe] (right-hand panel) for the simulated datasets compared with \emph{GES-iDR4}. 
As we have for Figure~\ref{agedf}, we also provide kernel density estimates of the [Fe/H] and [Mg/Fe] \emph{GES-iDR4} distributions in Figure~\ref{mdf} to show the effect of the 
uncertainties on these measurements. The uncertainties have a much smaller impact on these abundance ratios than on the age distribution but the peak in [Mg/Fe] is noticeably broadened. 
Again, we stress that the processed simulation data should be compared with \emph{GES-iDR4} and not \emph{GES-iDR4-KDE} which is simply shown to illustrate the 
effect of broadening in accordance with uncertainties.

The unaltered simulated stellar population particles (\emph{Selene-CH}, blue line) have a more peaked [Fe/H]} distribution compared to the observations (black line), and an extremely narrow distribution in [Mg/Fe]. Furthermore, the mass-weighted \emph{Selene-CH} distribution functions are truncated, at high-[Fe/H] and at both low- and high-[Mg/Fe]. Table~\ref{thetable} shows the interquartile range (IQR), skewness ($\sigma_{3}$), and kurtosis\footnote{We use the definition of kurtosis whereby a normal distribution has a kurtosis=0 (the excess kurtosis)} ($\sigma_{4}$), which allow us to perform a quantitative analysis of the effect of the observationally motivated changes to the simulated data (see below).

The differences between the width of the \emph{Selene-CH} distributions and the observations do not indicate a failure of the simulation. The simulated [Mg/Fe] 
distribution is smoothed and broadened significantly when the observationally-motivated scattering with errors is applied (\emph{Selene-GES}, green line). 
This is seen as the increase in the IQR for the [Mg/Fe], and to a lesser extent, in the [Fe/H] distributions. The effect is most pronounced for [Mg/Fe] where 
wings are created in the data on both sides of the distribution. For [Fe/H], the change is only noticeable for higher [Fe/H] values, with the low-metallicity tail 
being largely unaffected. The key result of Figure~\ref{mdf} is that the observational uncertainties in age, metallicity, and [Mg/Fe] have a much greater effect 
on the resulting distribution functions than do the photometric selection filters. This means that the observational uncertainties place a fundamental limit 
on detection of any substructure and on our ability to quantify the slope of any astrophysical relevant relationship in the data.

As expected, scattering the data leads to an increase in the IQR for both the [Fe/H] and [Mg/Fe] distributions providing a good agreement with the observed IQR compared to the original values. When followed up by imposing selection functions we find that the distributions are slightly narrowed but not so much that the reasonable agreement  in the spread of the distributions are lost. The simulated [Fe/H] distribution is improved by both stages of our post-processing with the distribution becoming less skewed and reducing in kurtosis to approach the observed values largely due to the enhanced positive tail of the distribution. The [Mg/Fe] distribution is slightly more complicated in that the post-processing does not give a particularly good qualitative fit to the observations in terms of skewness and kurtosis despite the success of reproducing the IQR. As with the [Fe/H] distribution the scattering and selection effects make the initially negatively-skewed distribution more positive but does not go far enough to be in line with the positively-skewed, \emph{GES-iDR4}, [Mg/Fe] data. The observed [Mg/Fe] kurtosis indicates a higher likelihood of outliers than found with a normal distribution, the even higher value of the \emph{Selene-CH} distribution is due to the extremely narrow distribution (kurtosis is not a measure of peakedness). The post-processing greatly reduces the kurtosis to be much closer to zero which is entirely expected as the scattering in particular pushes the distribution to be almost normal. The conformity of the \emph{Selene-SYN} distribution to a normal curve is because of the initially narrow distribution and the large scale of the scattering from the [Mg/Fe] uncertainty.

While the width of the observed DFs can be reproduced by application of observationally-motivated scattering, our simulations do not recover the detailed shape 
of the [Fe/H] or [Mg/Fe] distribution functions. The shape of the simulated [Fe/H] distribution is promisingly close but still defies similarity with an excess 
of stars between -0.4 and -0.2 and a deficit between -0.6 and -0.4; however the mean values of the [Fe/H] and [Mg/Fe] distributions do both match the observed data. The post-processing does 
not give a particularly good qualitative fit to the observations in terms of skewness and kurtosis. As with the [Fe/H] distribution the scattering and selection 
effects make the initially negatively-skewed distribution more positive but does not go far enough to be in line with the positively-skewed, \emph{GES-iDR4}, 
[Mg/Fe] data.

The mismatch between observed and simulated data for [Mg/Fe] could also hint at the problem with the observations or with stellar yields in our chemical evolution model. In fact, our results confirm the earlier studies \citep{1995ApJS...98..617T,2004A&A...421..613F,2017ApJ...835..224A} that show that chemical evolution models of the solar neighbourhood systematically under-predict [Mg/Fe] at any metallicity, also the solar values are too low compared to the observed [Mg/Fe] in the solar photosphere. This could be either due to poorly understood stellar yields of SNIa or SNII \citep[see][]{2004A&A...421..613F}, or because of the systematic errors in the observed data. It is known that Mg lines in cool stars are affected by NLTE \citep{2015ApJ...804..113B, 2011MNRAS.418..863M}. In fact, \citet[][submitted]{2016arXiv161207363B}, show that the NLTE [Mg/Fe] trend is lower than LTE trend, that would help to improve the agreement with the simulations.

\begin{table}
\centering
\begin{tabular}{ccccccc}
\hline\hline
     & \multicolumn{3}{c}{[Fe/H]} & \multicolumn{3}{c}{[Mg/Fe]} \\
  Name  & IQR & $\sigma_{3}$ & $\sigma_{4}$ & IQR & $\sigma_{3}$ & $\sigma_{4}$\\
\hline
  \emph{Selene-CH}  & 0.398 & -1.47 & 3.52 & 0.113 & -0.282 & 4.26\\
  \emph{Selene-GES} & 0.409 & -1.28 & 2.98 & 0.180 & -0.0165 & 0.358\\
  \emph{Selene-SYN} & 0.350 & -0.98 & 1.15 & 0.160 & 0.0416 & 0.099\\
\hline
  \emph{GES-iDR4}    & 0.414 & -0.63 & 0.95 & 0.175 & 1.04 & 3.64\\
\hline
\end{tabular}
\caption{[Fe/H] and [Mg/Fe] distribution function characteristics for the three simulation datasets (\emph{Selene-CH}, \emph{Selene-GES} and \emph{Selene-SYN}), 
and the observational dataset \emph{GES-iDR4} in dex. The interquartile range (IQR), skewness ($\sigma_{3}$), and kurtosis ($\sigma_{4}$) values for the 
distribution functions of [Fe/H] and [Mg/Fe] as shown in Figure~\ref{mdf}. Columns 1 gives the name of the dataset, columns 2, 3  
and 4 are the IQR, $\sigma_{3}$, and $\sigma_{4}$ of the [Fe/H] distribution. Columns 5, 6 and 7 are IQR, $\sigma_{3}$, and $\sigma_{4}$
of the [Mg/Fe] distribution.}
\label{thetable}
\end{table}

\subsection{Relationships between [Fe/H], [Mg/Fe], and stellar ages in the observed and simulated solar neighbourhood}

\begin{figure*}
\centering
\includegraphics[width=1.0\textwidth]{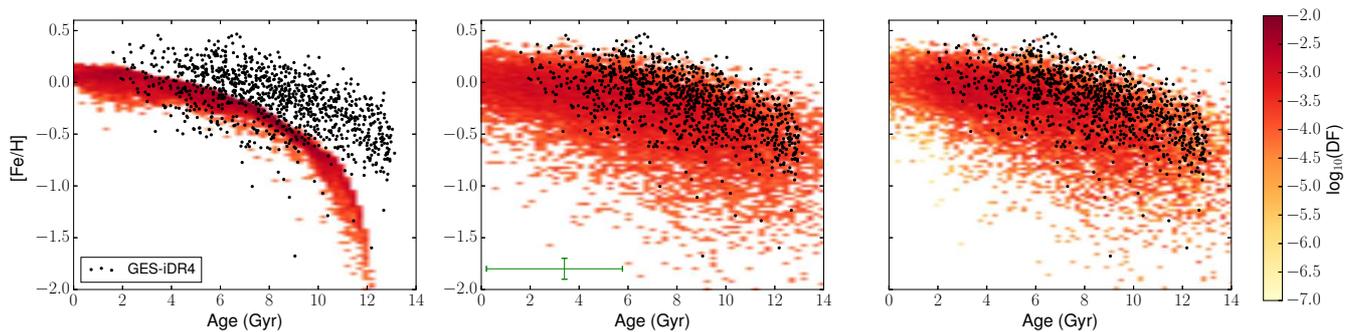}
\caption{Age-metallicity relation for our various simulated datasets compared 
with the \emph{GES-iDR4} distribution. The \emph{GES-iDR4} stars 
are plotted as black points in each panel. \emph{Selene-CH}, \emph{Selene-GES}
 and \emph{Selene-SYN} are represented as normalised heat maps with increasingly red colours indicating an 
increase in the abundance of composite or synthetic star particles in bins of 0.025 dex in 
[Fe/H] and 0.2 Gyr in age. we include a representative
error bar in green which represents the size of the scatter in [Fe/H] and age between \emph{Selene-CH} and
\emph{Selene-GES} ( $\sigma_{\mathrm{[Fe/H]}}$~=~0.101~dex, $\sigma_{age,l}$~=~3.20~Gyr and $\sigma_{age,u}$~=~2.37~Gyr, values which  are computed
from the mean of the errors of the \emph{GES-iDR4} data.).}
\label{agemet}
\end{figure*}

\begin{figure*}
\centering
\includegraphics[width=1.0\textwidth]{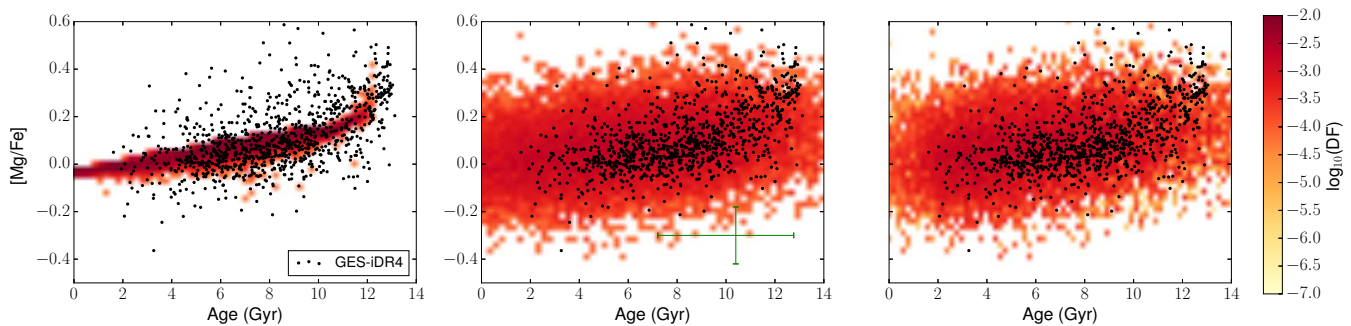}
\caption{[Mg/Fe] versus stellar age. The \emph{GES-iDR4} stars 
are plotted as black points in each panel. \emph{Selene-CH}, \emph{Selene-GES}
 and \emph{Selene-SYN} are represented as normalised heat maps with increasingly red colours indicating an 
increase in the abundance of composite or synthetic star particles in bins of 0.025 dex in 
[Fe/H] and 0.2 Gyr in age. we include a representative
error bar in green which represents the size of the scatter in [Mg/Fe] and age between \emph{Selene-CH} and
\emph{Selene-GES} ($\sigma_{\mathrm{[Mg/Fe]}}$~=~0.120~dex, 
$\sigma_{age,l}$~=~3.20~Gyr and $\sigma_{age,u}$~=~2.37~Gyr, values which are computed
from the mean of the errors of the \emph{GES-iDR4} data.).}
\label{agemg}
\end{figure*}

While the quantitative analysis of 1D distribution functions is precise, it does not aid our understanding of which stars are 
responsible for the differences between the models and simulations. Greater insight can be provided by examining the evolution 
of [Fe/H] and [Mg/Fe] with age which are shown in Figures~\ref{agemet} and \ref{agemg} respectively.

The distribution of the raw simulation data with only a geographical cut (\emph{Selene-CH}) in Figure~\ref{agemet} is significantly narrower 
than the observed distribution at all ages. The distribution of stellar population particles is particularly narrow for the oldest stars 
which are responsible for the low-metallicity tail in Figure~\ref{mdf}.
The narrow distributions of the \emph{Selene-CH} stellar population particles in both [Fe/H] (Figure~\ref{agemet}) and [Mg/Fe] (Figure~\ref{agemg}) are 
a result of the overall trend with age which contrasts with the observed distribution where observational uncertainties dominate. 
As discussed in the previous section, when the scatter is applied in \emph{Selene-GES}, the initially narrow underlying distribution 
is no longer discernible in the simulated data and the distribution is instead far more consistent with the observed data albeit with a 
slight offset to younger ages at a given [Fe/H].

The original, unscattered distribution of the stellar population particles underestimates [Fe/H] compared to the 
observed distribution for particles older than 8~Gyr (left-hand panel of Figure~\ref{agemet}).
This underestimation in the old stars is not so prominent in the \emph{Selene-GES} results 
(middle panel of Figure~\ref{agemet}), because scattering with observational age errors brings some of the young metal-rich stars to greater 
apparent ages. Applying the \textsc{SynCMD} tool to produce \emph{Selene-SYN} does not lead to any significant improvement 
over \emph{Selene-GES} in terms of fitting the observations. While some of the youngest stars are removed (see Figure~\ref{agedf}), 
it is not sufficient to match the dearth of young stars in \emph{GES-iDR4}. 

The trend of [Mg/Fe] with age for the unaltered stellar population particles (left-hand panel of Figure~\ref{agemg}) is a roughly 
linear increase with a very small up-turn seen in the oldest stars. Again the scattering broadens the distribution significantly 
(as shown in the middle panel of Figure~\ref{agemg}), but the application of the selection function makes only small changes to the 
distribution of the simulated stars (right-hand panel of Figure~\ref{agemg}). The main difference between simulation and observation 
is in the stars older than 10~Gyr; the up-turn in the underlying \emph{Selene-CH} is not as strong as for the observed old and high-[Mg/Fe] stars. 
As discussed earlier, there are two possible explanations: a) the neglected NLTE effects in our observed [Mg/Fe] distributions, 
and b) erroneous stellar yields in the chemical evolution model.

Finally, we should note that the GES high-resolution dataset does not show any evidence of the bimodal [Mg/Fe] distribution with [Fe/H], which have been proposed as a chemical separator of the thin and thick discs. 
This is also consistent with our simulations, which do not have any discontinuity in the SFH at $\sim$1 Gyr.

\section{Conclusions}
\label{conclusion}

We compare the results of a Milky Way-like galaxy simulation created using \textsc{ramses-ch} \citep{2014MNRAS.444.3845F} with 
the fourth data release of the Gaia-ESO survey considering the 1D distribution functions of age, [Fe/H], and [Mg/Fe] as well as the  
age evolution of the latter two properties. The comparison is conducted in three stages; 
\begin{enumerate}
\item The simulated stellar population particles are compared directly with the observed distributions.
\item Typical observational uncertainty (from \emph{GES-iDR4}) as the standard deviation of a Gaussian function used to stochastically scatter the simulated data to mimic observational uncertainty.
\item The simulated stellar population particles are stochastically scattered as above and are then split into individual stars based on stellar population models and only those accepted by the \emph{GES-iDR4} selection functions are retained for comparison.
\end{enumerate}

Each of these stages mimics the effects found in observations as a way of placing the simulated data in the `observer frame'.
The application of stochastic scattering based on the errors of an observational survey has the effect of smoothing out the age distribution function. The further application of the \emph{GES-iDR4} selection function has the effect of removing young stars  (ages~<~1~Gyr) and shifting the peak age to between 2 and 5~Gyr. Despite both these effects bringing the simulated age distribution closer to the observed one there is still a significant offset between the distributions. One possible explanation to the remaining discrepancies is that our simulated galaxy has a different assembly history. \emph{Selene} (original simulation from \citet{2012A&A...547A..63F}. is not constrained to be identical to the Milky Way and thus determining the location in \emph{Selene-CH} which is the best analogue to the solar neighbourhood is somewhat open to interpretation.

Our key finding is that there is no need for the chemical evolution models to reproduce the large dispersion in metallicity and age seen in observational studies \citep[e.g][]{2014A&A...565A..89B}. We show here that this dispersion is apparent and can be fully accounted for by the observational uncertainty. The typical observational uncertainties of our dataset are of the order 30\% for age and 0.1 dex for [Fe/H] and [Mg/Fe]. When we apply errors to our simulated stars, we get the age-[Fe/H]-[Mg/Fe] distribution functions that are more consistent with the observed data. In particular, observationally-motivated scattering spreads the [Fe/H] distribution towards larger values, which creates an apparent high-metallicity tail. This leads to a larger fraction of metal-rich stars at all ages, improving the agreement with the observed metallicity distribution for the oldest stars.

The scattering and observational selection function improve not only the fit in the age-abundance and age-metallicity plane, but also the fit of the of the simulated [Fe/H] and [Mg/Fe] distribution functions to the \emph{GES-iDR4} data. Scattering according to observational uncertainties broadens the distribution functions (increased IQR) and swamps their excessive negative skewness, driving them toward a normal curve. The scattering also produces wings on both sides of the initially narrow simulated [Mg/Fe] distribution, making it more consistent with observations.
The application of an observational selection function to the simulated data acts to slightly reduce the IQR by culling a number of outliers from the distribution tails. This effect, in particular, removes most metal-rich and metal-poor stars, which are most likely to be the oldest and youngest stars. As a consequence, the kurtosis of both [Fe/H] and [Mg/Fe] distribution functions is reduced. Also our normal scattering function drives the distributions to conform more closely to a normal curve.

One peculiar feature of the observed [Mg/Fe] distributions is that they have a positive skew, which is very difficult to induce in the simulations. It is possible that our assumption of Gaussian errors is wrong. However, at present we do not have a full error matrix for Gaia-ESO dataset  and cannot account for more complex correlations and systematic effects in the observational data.

We also stress that neither in the data, nor in the simulations do we detect the bimodality in the [Mg/Fe] space with metallicity, which has been claimed by some studies \citep[e.g][]{1998A&A...338..161F} to trace the Galactic thin and thick discs.  The assembly history of the simulated galaxy is relatively quiescent with no major mergers after redshift z $= 1.0$, and there are no prominent sub-structures in the simulated [Fe/H] - [Mg/Fe] plane, which could result from an episodic star formation history \citep[see][]{2014MNRAS.444.3845F}.

Our analysis also reveals that there is more fundamental mismatch between the models and observations: the high-[Mg/Fe] stars generally lack in the original and scattered simulated datasets. This problem most likely has a different cause. Firstly, it has been demonstrated that the NLTE corrections to Mg abundances for the late F- and GK-type stars at lower metallicity, [Fe/H]~$\sim$-0.5~to~-1.5~dex, are usually negative \citep{2016A&A...586A.120O,2016arXiv161207363B}, that is the LTE abundances of Mg are overestimated. Therefore, the high-[Mg/Fe] plateau could be a consequence of the systematics in the observed LTE dataset. Secondly, the stellar yields of Mg and/or Fe are not well understood. Standard chemical evolution models severely underproduce the solar Mg abundance \citep{2017ApJ...835..224A}, but also the shape of the [Mg/Fe] with metallicity does not conform to observations. While this problem is beyond the scope of our paper we would like to encourage future work to explore the causes of this phenomena.

To conclude, it is fundamentally important to reduce the uncertainty of the observed datasets in order to constrain the models of Galaxy formation. The typical observational uncertainty of 0.1 dex in chemical abundances and $\sim$$30\%$ in ages are too large to provide meaningful information on the substructure in the age-chemical abundance space, which is relevant to the interpretation of the evolution of the Galactic disc. Survey selection functions, like the colour-magnitude selection in the Gaia-ESO survey, may or may not have a sizeable effect on the results, however for the Gaia-ESO, this effect is extremely small compared to the effect of observational uncertainties. 

\section*{Acknowledgements}
We acknowledge the insightful comments and support provided by our 
colleagues Stefano Pasetto, Daisuke Kawata, Rob Thacker and Dimitris Stamatellos. 
We would thank the anonymous referee for
a very constructive report of the work presented here.
B.B.T. acknowledges the support of STFC through its PhD Studentship programme 
(ST/F007701/1). 
We also acknowledge the generous allocation of resources 
from the Partnership for Advanced Computing in Europe (PRACE) via the 
DEISA Extreme Computing Initiative (PRACE-3IP Project RI-312763 and 
PRACE-4IP Project 653838), and STFC's DiRAC Facility (COSMOS: Galactic 
Archaeology). C.G.F acknowledges funding from the European Research Council for the FP7 ERC starting grant
project LOCALSTAR and the DiRAC Complexity system, operated by the University of Leicester IT
Services, which forms part of the STFC DiRAC HPC Facility (www.dirac.ac.uk). This equipment is funded by BIS
National E-Infrastructure capital grant ST/K000373/1 and
STFC DiRAC Operations grant ST/K0003259/1. DiRAC is part of the National E-Infrastructure. 
Continued access to the University of Hull's High 
Performance Computing Facility (`viper'),
the HPC facility at the University of Central Lancashire and the computational
facilities at Saint Mary's University are 
likewise gratefully acknowledged. T.B. was funded by the project grant 
``The New Milky Way''  from the Knut and Alice Wallenberg Foundation. 
S.G.S acknowledges the support by Funda\c{c}\~{a}o para a Ci\^{e}ncia e Tecnologia 
(FCT) (ref: UID/FIS/04434/2013 \& PTDC/FIS-AST/7073/2014 \& 
Investigador FCT contract of reference IF/00028/2014) through national
 funds and by FEDER through COMPETE2020
(ref: POCI-01-0145-FEDER-007672 \& POCI-01-0145-FEDER-016880).
U.H. acknowledges support from the Swedish National Space Board (SNSB/Rymdstyrelsen).
The Gaia-ESO Survey data products have been processed by the Cambridge Astronomy Survey Unit (CASU) 
at the Institute of Astronomy, University of Cambridge, and by the FLAMES/UVES 
reduction team at INAF/Osservatorio Astrofisico di Arcetri. 
These data have been obtained from the Gaia-ESO Survey Data Archive,
 prepared and hosted by the Wide Field Astronomy Unit,
 Institute for Astronomy, University of Edinburgh, which is funded 
by the UK Science and Technology Facilities Council.
This work was partly supported by the European Union FP7 programme 
through ERC grant number 320360 and by the Leverhulme Trust through grant RPG-2012-541. 
We acknowledge the support from INAF and Ministero dell' Istruzione, dell' 
Universit\`a' e della Ricerca (MIUR) in the form of the grant ``Premiale VLT 2012''.
 The results presented here benefit from discussions held during the Gaia-ESO workshops 
and conferences supported by the ESF (European Science Foundation) through the GREAT 
Research Network Programme.
M.T.C acknowledge the financial support from the Spanish Ministerio de Econom\'{i}a y Competitividad,
 through grant AYA2013-40611-P. U.H. acknowledges support from the Swedish National Space Board (SNSB/Rymdstyrelsen).
This work was supported by Sonderforschungsbereich SFB 881 ``The Milky Way System'' 
(subprojects A5, C9) of the German Research Foundation (DFG).
This work benefited from discussions at GNASH workshop, Victoria supported by the National Science Foundation under Grant No. PHY-1430152 (JINA Center for the Evolution of the Elements).

\bibliography{gaiaeso}{}

\begin{thebibliography}{}
\makeatletter
\relax
\def\mn@urlcharsother{\let\do\@makeother \do\$\do\&\do\#\do\^\do\_\do\%\do\~}
\def\mn@doi{\begingroup\mn@urlcharsother \@ifnextchar [ {\mn@doi@}
  {\mn@doi@[]}}
\def\mn@doi@[#1]#2{\def\@tempa{#1}\ifx\@tempa\@empty \href
  {http://dx.doi.org/#2} {doi:#2}\else \href {http://dx.doi.org/#2} {#1}\fi
  \endgroup}
\def\mn@eprint#1#2{\mn@eprint@#1:#2::\@nil}
\def\mn@eprint@arXiv#1{\href {http://arxiv.org/abs/#1} {{\tt arXiv:#1}}}
\def\mn@eprint@dblp#1{\href {http://dblp.uni-trier.de/rec/bibtex/#1.xml}
  {dblp:#1}}
\def\mn@eprint@#1:#2:#3:#4\@nil{\def\@tempa {#1}\def\@tempb {#2}\def\@tempc
  {#3}\ifx \@tempc \@empty \let \@tempc \@tempb \let \@tempb \@tempa \fi \ifx
  \@tempb \@empty \def\@tempb {arXiv}\fi \@ifundefined
  {mn@eprint@\@tempb}{\@tempb:\@tempc}{\expandafter \expandafter \csname
  mn@eprint@\@tempb\endcsname \expandafter{\@tempc}}}

\bibitem[\protect\citeauthoryear{{Agertz} et~al.,}{{Agertz}
  et~al.}{2007}]{2007MNRAS.380..963A}
{Agertz} O.,  et~al., 2007, \mn@doi [\mnras]
  {10.1111/j.1365-2966.2007.12183.x}, \href
  {http://adsabs.harvard.edu/abs/2007MNRAS.380..963A} {380, 963}

\bibitem[\protect\citeauthoryear{{Andrews}, {Weinberg}, {Sch{\"o}nrich}  \&
  {Johnson}}{{Andrews} et~al.}{2017}]{2017ApJ...835..224A}
{Andrews} B.~H.,  {Weinberg} D.~H.,  {Sch{\"o}nrich} R.,   {Johnson} J.~A.,
  2017, \mn@doi [\apj] {10.3847/1538-4357/835/2/224}, \href
  {http://adsabs.harvard.edu/abs/2017ApJ...835..224A} {835, 224}

\bibitem[\protect\citeauthoryear{{Asplund}, {Grevesse}, {Sauval}  \&
  {Scott}}{{Asplund} et~al.}{2009}]{2009ARA&A..47..481A}
{Asplund} M.,  {Grevesse} N.,  {Sauval} A.~J.,   {Scott} P.,  2009, \mn@doi
  [\araa] {10.1146/annurev.astro.46.060407.145222}, \href
  {http://adsabs.harvard.edu/abs/2009ARA%26A..47..481A} {47, 481}

\bibitem[\protect\citeauthoryear{{Bergemann} et~al.,}{{Bergemann}
  et~al.}{2014}]{2014A&A...565A..89B}
{Bergemann} M.,  et~al., 2014, \mn@doi [\aap] {10.1051/0004-6361/201423456},
  \href {http://adsabs.harvard.edu/abs/2014A%26A...565A..89B} {565, A89}

\bibitem[\protect\citeauthoryear{{Bergemann}, {Kudritzki}, {Gazak}, {Davies}
  \& {Plez}}{{Bergemann} et~al.}{2015}]{2015ApJ...804..113B}
{Bergemann} M.,  {Kudritzki} R.-P.,  {Gazak} Z.,  {Davies} B.,   {Plez} B.,
  2015, \mn@doi [\apj] {10.1088/0004-637X/804/2/113}, \href
  {http://adsabs.harvard.edu/abs/2015ApJ...804..113B} {804, 113}

\bibitem[\protect\citeauthoryear{{Bergemann}, {Collet}, {Schoenrich}, {Andrae},
  {Kovalev}, {Ruchti}, {Hansen}  \& {Magic}}{{Bergemann}
  et~al.}{2016}]{2016arXiv161207363B}
{Bergemann} M.,  {Collet} R.,  {Schoenrich} R.,  {Andrae} R.,  {Kovalev} M.,
  {Ruchti} G.,  {Hansen} C.~J.,   {Magic} Z.,  2016, ApJ in press, \href
  {http://esoads.eso.org/abs/2016arXiv161207363B} {}

\bibitem[\protect\citeauthoryear{{Bertelli}, {Nasi}, {Girardi}, {Chiosi},
  {Zoccali}  \& {Gallart}}{{Bertelli} et~al.}{2003}]{2003AJ....125..770B}
{Bertelli} G.,  {Nasi} E.,  {Girardi} L.,  {Chiosi} C.,  {Zoccali} M.,
  {Gallart} C.,  2003, \mn@doi [\aj] {10.1086/345961}, \href
  {http://adsabs.harvard.edu/abs/2003AJ....125..770B} {125, 770}

\bibitem[\protect\citeauthoryear{{Bertelli}, {Girardi}, {Marigo}  \&
  {Nasi}}{{Bertelli} et~al.}{2008}]{2008A&A...484..815B}
{Bertelli} G.,  {Girardi} L.,  {Marigo} P.,   {Nasi} E.,  2008, \mn@doi [\aap]
  {10.1051/0004-6361:20079165}, \href
  {http://cdsads.u-strasbg.fr/abs/2008A%26A...484..815B} {484, 815}

\bibitem[\protect\citeauthoryear{{Bertelli}, {Nasi}, {Girardi}  \&
  {Marigo}}{{Bertelli} et~al.}{2009}]{2009A&A...508..355B}
{Bertelli} G.,  {Nasi} E.,  {Girardi} L.,   {Marigo} P.,  2009, \mn@doi [\aap]
  {10.1051/0004-6361/200912093}, \href
  {http://cdsads.u-strasbg.fr/abs/2009A%26A...508..355B} {508, 355}

\bibitem[\protect\citeauthoryear{{Calura} \& {Menci}}{{Calura} \&
  {Menci}}{2009}]{2009MNRAS.400.1347C}
{Calura} F.,  {Menci} N.,  2009, \mn@doi [\mnras]
  {10.1111/j.1365-2966.2009.15440.x}, \href
  {http://adsabs.harvard.edu/abs/2009MNRAS.400.1347C} {400, 1347}

\bibitem[\protect\citeauthoryear{{Calura} et~al.,}{{Calura}
  et~al.}{2012}]{2012MNRAS.427.1401C}
{Calura} F.,  et~al., 2012, \mn@doi [\mnras]
  {10.1111/j.1365-2966.2012.22052.x}, \href
  {http://adsabs.harvard.edu/abs/2012MNRAS.427.1401C} {427, 1401}

\bibitem[\protect\citeauthoryear{{Chabrier}}{{Chabrier}}{2003}]{2003ApJ...586L.133C}
{Chabrier} G.,  2003, \mn@doi [\apjl] {10.1086/374879}, \href
  {http://adsabs.harvard.edu/abs/2003ApJ...586L.133C} {586, L133}

\bibitem[\protect\citeauthoryear{{Chiappini}, {Matteucci}  \&
  {Romano}}{{Chiappini} et~al.}{2001}]{2001ApJ...554.1044C}
{Chiappini} C.,  {Matteucci} F.,   {Romano} D.,  2001, \mn@doi [\apj]
  {10.1086/321427}, \href {http://adsabs.harvard.edu/abs/2001ApJ...554.1044C}
  {554, 1044}

\bibitem[\protect\citeauthoryear{{Dekker}, {D'Odorico}, {Kaufer}, {Delabre}  \&
  {Kotzlowski}}{{Dekker} et~al.}{2000}]{2000SPIE.4008..534D}
{Dekker} H.,  {D'Odorico} S.,  {Kaufer} A.,  {Delabre} B.,   {Kotzlowski} H.,
  2000, in {Iye} M.,  {Moorwood} A.~F.,  eds,  \procspie Vol. 4008, Optical and
  IR Telescope Instrumentation and Detectors. pp 534--545,
  \mn@doi{10.1117/12.395512}

\bibitem[\protect\citeauthoryear{{Dobbs} et~al.,}{{Dobbs}
  et~al.}{2017}]{2017MNRAS.464.3580D}
{Dobbs} C.~L.,  et~al., 2017, \mn@doi [\mnras] {10.1093/mnras/stw2200}, \href
  {http://adsabs.harvard.edu/abs/2017MNRAS.464.3580D} {464, 3580}

\bibitem[\protect\citeauthoryear{{Eggen}, {Lynden-Bell}  \& {Sandage}}{{Eggen}
  et~al.}{1962}]{1962ApJ...136..748E}
{Eggen} O.~J.,  {Lynden-Bell} D.,   {Sandage} A.~R.,  1962, \mn@doi [\apj]
  {10.1086/147433}, \href {http://adsabs.harvard.edu/abs/1962ApJ...136..748E}
  {136, 748}

\bibitem[\protect\citeauthoryear{{Ferland}, {Korista}, {Verner}, {Ferguson},
  {Kingdon}  \& {Verner}}{{Ferland} et~al.}{1998}]{1998PASP..110..761F}
{Ferland} G.~J.,  {Korista} K.~T.,  {Verner} D.~A.,  {Ferguson} J.~W.,
  {Kingdon} J.~B.,   {Verner} E.~M.,  1998, \mn@doi [\pasp] {10.1086/316190},
  \href {http://adsabs.harvard.edu/abs/1998PASP..110..761F} {110, 761}

\bibitem[\protect\citeauthoryear{{Few}, {Courty}, {Gibson}, {Kawata}, {Calura}
  \& {Teyssier}}{{Few} et~al.}{2012a}]{2012MNRAS.424L..11F}
{Few} C.~G.,  {Courty} S.,  {Gibson} B.~K.,  {Kawata} D.,  {Calura} F.,
  {Teyssier} R.,  2012a, \mn@doi [\mnras] {10.1111/j.1745-3933.2012.01275.x},
  \href {http://adsabs.harvard.edu/abs/2012MNRAS.424L..11F} {424, L11}

\bibitem[\protect\citeauthoryear{{Few}, {Gibson}, {Courty}, {Michel-Dansac},
  {Brook}  \& {Stinson}}{{Few} et~al.}{2012b}]{2012A&A...547A..63F}
{Few} C.~G.,  {Gibson} B.~K.,  {Courty} S.,  {Michel-Dansac} L.,  {Brook}
  C.~B.,   {Stinson} G.~S.,  2012b, \mn@doi [\aap]
  {10.1051/0004-6361/201219649}, \href
  {http://adsabs.harvard.edu/abs/2012A%26A...547A..63F} {547, A63}

\bibitem[\protect\citeauthoryear{{Few}, {Courty}, {Gibson}, {Michel-Dansac}  \&
  {Calura}}{{Few} et~al.}{2014}]{2014MNRAS.444.3845F}
{Few} C.~G.,  {Courty} S.,  {Gibson} B.~K.,  {Michel-Dansac} L.,   {Calura} F.,
   2014, \mn@doi [\mnras] {10.1093/mnras/stu1709}, \href
  {http://adsabs.harvard.edu/abs/2014MNRAS.444.3845F} {444, 3845}

\bibitem[\protect\citeauthoryear{{Few}, {Dobbs}, {Pettitt}  \&
  {Konstandin}}{{Few} et~al.}{2016}]{2016MNRAS.460.4382F}
{Few} C.~G.,  {Dobbs} C.,  {Pettitt} A.,   {Konstandin} L.,  2016, \mn@doi
  [\mnras] {10.1093/mnras/stw1226}, \href
  {http://adsabs.harvard.edu/abs/2016MNRAS.460.4382F} {460, 4382}

\bibitem[\protect\citeauthoryear{{Fran{\c c}ois}, {Matteucci}, {Cayrel},
  {Spite}, {Spite}  \& {Chiappini}}{{Fran{\c c}ois}
  et~al.}{2004}]{2004A&A...421..613F}
{Fran{\c c}ois} P.,  {Matteucci} F.,  {Cayrel} R.,  {Spite} M.,  {Spite} F.,
  {Chiappini} C.,  2004, \mn@doi [\aap] {10.1051/0004-6361:20034140}, \href
  {http://cdsads.u-strasbg.fr/abs/2004A%26A...421..613F} {421, 613}

\bibitem[\protect\citeauthoryear{{Freeman} \& {Bland-Hawthorn}}{{Freeman} \&
  {Bland-Hawthorn}}{2002}]{2002ARA&A..40..487F}
{Freeman} K.,  {Bland-Hawthorn} J.,  2002, \mn@doi [\araa]
  {10.1146/annurev.astro.40.060401.093840}, \href
  {http://adsabs.harvard.edu/abs/2002ARA%26A..40..487F} {40, 487}

\bibitem[\protect\citeauthoryear{{Fuhrmann}}{{Fuhrmann}}{1998}]{1998A&A...338..161F}
{Fuhrmann} K.,  1998, \aap, \href
  {http://adsabs.harvard.edu/abs/1998A%26A...338..161F} {338, 161}

\bibitem[\protect\citeauthoryear{{Gibson}, {Pilkington}, {Brook}, {Stinson}  \&
  {Bailin}}{{Gibson} et~al.}{2013}]{2013A&A...554A..47G}
{Gibson} B.~K.,  {Pilkington} K.,  {Brook} C.~B.,  {Stinson} G.~S.,   {Bailin}
  J.,  2013, \mn@doi [\aap] {10.1051/0004-6361/201321239}, \href
  {http://cdsads.u-strasbg.fr/abs/2013A%26A...554A..47G} {554, A47}

\bibitem[\protect\citeauthoryear{{Gill}, {Knebe}  \& {Gibson}}{{Gill}
  et~al.}{2004}]{2004MNRAS.351..399G}
{Gill} S.~P.~D.,  {Knebe} A.,   {Gibson} B.~K.,  2004, \mn@doi [\mnras]
  {10.1111/j.1365-2966.2004.07786.x}, \href
  {http://adsabs.harvard.edu/abs/2004MNRAS.351..399G} {351, 399}

\bibitem[\protect\citeauthoryear{{Gilmore} et~al.,}{{Gilmore}
  et~al.}{2012}]{2012Msngr.147...25G}
{Gilmore} G.,  et~al., 2012, The Messenger, \href
  {http://adsabs.harvard.edu/abs/2012Msngr.147...25G} {147, 25}

\bibitem[\protect\citeauthoryear{{Gustafsson}, {Edvardsson}, {Eriksson},
  {J{\o}rgensen}, {Nordlund}  \& {Plez}}{{Gustafsson}
  et~al.}{2008}]{2008A&A...486..951G}
{Gustafsson} B.,  {Edvardsson} B.,  {Eriksson} K.,  {J{\o}rgensen} U.~G.,
  {Nordlund} {\AA}.,   {Plez} B.,  2008, \mn@doi [\aap]
  {10.1051/0004-6361:200809724}, \href
  {http://adsabs.harvard.edu/abs/2008A%26A...486..951G} {486, 951}

\bibitem[\protect\citeauthoryear{{Haardt} \& {Madau}}{{Haardt} \&
  {Madau}}{1996}]{1996ApJ...461...20H}
{Haardt} F.,  {Madau} P.,  1996, \mn@doi [\apj] {10.1086/177035}, \href
  {http://adsabs.harvard.edu/abs/1996ApJ...461...20H} {461, 20}

\bibitem[\protect\citeauthoryear{{Heiter} et~al.,}{{Heiter}
  et~al.}{2015}]{2015PhyS...90e4010H}
{Heiter} U.,  et~al., 2015, \mn@doi [\physscr] {10.1088/0031-8949/90/5/054010},
  \href {http://adsabs.harvard.edu/abs/2015PhyS...90e4010H} {90, 054010}

\bibitem[\protect\citeauthoryear{{Henry}, {Kwitter}, {Jaskot}, {Balick},
  {Morrison}  \& {Milingo}}{{Henry} et~al.}{2010}]{2010ApJ...724..748H}
{Henry} R.~B.~C.,  {Kwitter} K.~B.,  {Jaskot} A.~E.,  {Balick} B.,  {Morrison}
  M.~A.,   {Milingo} J.~B.,  2010, \mn@doi [\apj]
  {10.1088/0004-637X/724/1/748}, \href
  {http://cdsads.u-strasbg.fr/abs/2010ApJ...724..748H} {724, 748}

\bibitem[\protect\citeauthoryear{{Hopkins}}{{Hopkins}}{2015}]{2015MNRAS.450...53H}
{Hopkins} P.~F.,  2015, \mn@doi [\mnras] {10.1093/mnras/stv195}, \href
  {http://adsabs.harvard.edu/abs/2015MNRAS.450...53H} {450, 53}

\bibitem[\protect\citeauthoryear{{Huchra} et~al.,}{{Huchra}
  et~al.}{2012}]{2012ApJS..199...26H}
{Huchra} J.~P.,  et~al., 2012, \mn@doi [\apjs] {10.1088/0067-0049/199/2/26},
  \href {http://adsabs.harvard.edu/abs/2012ApJS..199...26H} {199, 26}

\bibitem[\protect\citeauthoryear{{Jofr{\'e}} et~al.,}{{Jofr{\'e}}
  et~al.}{2015}]{2015A&A...582A..81J}
{Jofr{\'e}} P.,  et~al., 2015, \mn@doi [\aap] {10.1051/0004-6361/201526604},
  \href {http://adsabs.harvard.edu/abs/2015A%26A...582A..81J} {582, A81}

\bibitem[\protect\citeauthoryear{{Kawata} \& {Gibson}}{{Kawata} \&
  {Gibson}}{2003}]{2003MNRAS.340..908K}
{Kawata} D.,  {Gibson} B.~K.,  2003, \mn@doi [\mnras]
  {10.1046/j.1365-8711.2003.06356.x}, \href
  {http://adsabs.harvard.edu/abs/2003MNRAS.340..908K} {340, 908}

\bibitem[\protect\citeauthoryear{{Kennicutt}}{{Kennicutt}}{1998}]{1998ApJ...498..541K}
{Kennicutt} Jr. R.~C.,  1998, \mn@doi [\apj] {10.1086/305588}, \href
  {http://adsabs.harvard.edu/abs/1998ApJ...498..541K} {498, 541}

\bibitem[\protect\citeauthoryear{{Kim} et~al.,}{{Kim}
  et~al.}{2014}]{2014ApJS..210...14K}
{Kim} J.-h.,  et~al., 2014, \mn@doi [\apjs] {10.1088/0067-0049/210/1/14}, \href
  {http://adsabs.harvard.edu/abs/2014ApJS..210...14K} {210, 14}

\bibitem[\protect\citeauthoryear{{Knollmann} \& {Knebe}}{{Knollmann} \&
  {Knebe}}{2009}]{2009ApJS..182..608K}
{Knollmann} S.~R.,  {Knebe} A.,  2009, \mn@doi [\apjs]
  {10.1088/0067-0049/182/2/608}, \href
  {http://adsabs.harvard.edu/abs/2009ApJS..182..608K} {182, 608}

\bibitem[\protect\citeauthoryear{{Kobayashi} \& {Nakasato}}{{Kobayashi} \&
  {Nakasato}}{2011}]{2011ApJ...729...16K}
{Kobayashi} C.,  {Nakasato} N.,  2011, \mn@doi [\apj]
  {10.1088/0004-637X/729/1/16}, \href
  {http://adsabs.harvard.edu/abs/2011ApJ...729...16K} {729, 16}

\bibitem[\protect\citeauthoryear{{Kodama} \& {Arimoto}}{{Kodama} \&
  {Arimoto}}{1997}]{1997A&A...320...41K}
{Kodama} T.,  {Arimoto} N.,  1997, \aap, \href
  {http://adsabs.harvard.edu/abs/1997A%26A...320...41K} {320, 41}

\bibitem[\protect\citeauthoryear{{Komatsu} et~al.,}{{Komatsu}
  et~al.}{2011}]{2011ApJS..192...18K}
{Komatsu} E.,  et~al., 2011, \mn@doi [\apjs] {10.1088/0067-0049/192/2/18},
  \href {http://adsabs.harvard.edu/abs/2011ApJS..192...18K} {192, 18}

\bibitem[\protect\citeauthoryear{{Kroupa}}{{Kroupa}}{2001}]{2001MNRAS.322..231K}
{Kroupa} P.,  2001, \mn@doi [\mnras] {10.1046/j.1365-8711.2001.04022.x}, \href
  {http://adsabs.harvard.edu/abs/2001MNRAS.322..231K} {322, 231}

\bibitem[\protect\citeauthoryear{{Kroupa}, {Tout}  \& {Gilmore}}{{Kroupa}
  et~al.}{1993}]{1993MNRAS.262..545K}
{Kroupa} P.,  {Tout} C.~A.,   {Gilmore} G.,  1993, \mn@doi [\mnras]
  {10.1093/mnras/262.3.545}, \href
  {http://adsabs.harvard.edu/abs/1993MNRAS.262..545K} {262, 545}

\bibitem[\protect\citeauthoryear{{Lia}, {Portinari}  \& {Carraro}}{{Lia}
  et~al.}{2002}]{2002MNRAS.330..821L}
{Lia} C.,  {Portinari} L.,   {Carraro} G.,  2002, \mn@doi [\mnras]
  {10.1046/j.1365-8711.2002.05118.x}, \href
  {http://adsabs.harvard.edu/abs/2002MNRAS.330..821L} {330, 821}

\bibitem[\protect\citeauthoryear{{Marigo} \& {Girardi}}{{Marigo} \&
  {Girardi}}{2007}]{2007A&A...469..239M}
{Marigo} P.,  {Girardi} L.,  2007, \mn@doi [\aap] {10.1051/0004-6361:20066772},
  \href {http://cdsads.u-strasbg.fr/abs/2007A%26A...469..239M} {469, 239}

\bibitem[\protect\citeauthoryear{{Mart{\'{\i}}nez-Serrano}, {Serna},
  {Dom{\'{\i}}nguez-Tenreiro}  \& {Moll{\'a}}}{{Mart{\'{\i}}nez-Serrano}
  et~al.}{2008}]{2008MNRAS.388...39M}
{Mart{\'{\i}}nez-Serrano} F.~J.,  {Serna} A.,  {Dom{\'{\i}}nguez-Tenreiro} R.,
   {Moll{\'a}} M.,  2008, \mn@doi [\mnras] {10.1111/j.1365-2966.2008.13383.x},
  \href {http://adsabs.harvard.edu/abs/2008MNRAS.388...39M} {388, 39}

\bibitem[\protect\citeauthoryear{{Merle}, {Th{\'e}venin}, {Pichon}  \&
  {Bigot}}{{Merle} et~al.}{2011}]{2011MNRAS.418..863M}
{Merle} T.,  {Th{\'e}venin} F.,  {Pichon} B.,   {Bigot} L.,  2011, \mn@doi
  [\mnras] {10.1111/j.1365-2966.2011.19540.x}, \href
  {http://adsabs.harvard.edu/abs/2011MNRAS.418..863M} {418, 863}

\bibitem[\protect\citeauthoryear{{Miranda}, {Macfarlane}  \&
  {Gibson}}{{Miranda} et~al.}{2014}]{2014nic..confE.149M}
{Miranda} M.~S.,  {Macfarlane} B.~A.,   {Gibson} B.~K.,  2014, in Proceedings
  of XIII Nuclei in the Cosmos (NIC XIII). p.~149 (\mn@eprint {arXiv}
  {1502.00444})

\bibitem[\protect\citeauthoryear{{Miranda} et~al.,}{{Miranda}
  et~al.}{2016}]{2016A&A...587A..10M}
{Miranda} M.~S.,  et~al., 2016, \mn@doi [\aap] {10.1051/0004-6361/201525789},
  \href {http://adsabs.harvard.edu/abs/2016A%26A...587A..10M} {587, A10}

\bibitem[\protect\citeauthoryear{{Oppenheimer} \& {Dav{\'e}}}{{Oppenheimer} \&
  {Dav{\'e}}}{2008}]{2008MNRAS.387..577O}
{Oppenheimer} B.~D.,  {Dav{\'e}} R.,  2008, \mn@doi [\mnras]
  {10.1111/j.1365-2966.2008.13280.x}, \href
  {http://adsabs.harvard.edu/abs/2008MNRAS.387..577O} {387, 577}

\bibitem[\protect\citeauthoryear{{Osorio} \& {Barklem}}{{Osorio} \&
  {Barklem}}{2016}]{2016A&A...586A.120O}
{Osorio} Y.,  {Barklem} P.~S.,  2016, \mn@doi [\aap]
  {10.1051/0004-6361/201526958}, \href
  {http://adsabs.harvard.edu/abs/2016A%26A...586A.120O} {586, A120}

\bibitem[\protect\citeauthoryear{{Pagel} \& {Tautvaisiene}}{{Pagel} \&
  {Tautvaisiene}}{1995}]{1995MNRAS.276..505P}
{Pagel} B.~E.~J.,  {Tautvaisiene} G.,  1995, \mn@doi [\mnras]
  {10.1093/mnras/276.2.505}, \href
  {http://adsabs.harvard.edu/abs/1995MNRAS.276..505P} {276, 505}

\bibitem[\protect\citeauthoryear{{Pasetto}, {Chiosi}  \& {Kawata}}{{Pasetto}
  et~al.}{2012}]{2012A&A...545A..14P}
{Pasetto} S.,  {Chiosi} C.,   {Kawata} D.,  2012, \mn@doi [\aap]
  {10.1051/0004-6361/201219698}, \href
  {http://adsabs.harvard.edu/abs/2012A%26A...545A..14P} {545, A14}

\bibitem[\protect\citeauthoryear{{Pilkington} et~al.,}{{Pilkington}
  et~al.}{2012}]{2012A&A...540A..56P}
{Pilkington} K.,  et~al., 2012, \mn@doi [\aap] {10.1051/0004-6361/201117466},
  \href {http://adsabs.harvard.edu/abs/2012A%26A...540A..56P} {540, A56}

\bibitem[\protect\citeauthoryear{{Price}}{{Price}}{2008}]{2008JCoPh.22710040P}
{Price} D.~J.,  2008, \mn@doi [Journal of Computational Physics]
  {10.1016/j.jcp.2008.08.011}, \href
  {http://adsabs.harvard.edu/abs/2008JCoPh.22710040P} {227, 10040}

\bibitem[\protect\citeauthoryear{{Randich}, {Gilmore}  \& {Gaia-ESO
  Consortium}}{{Randich} et~al.}{2013}]{2013Msngr.154...47R}
{Randich} S.,  {Gilmore} G.,   {Gaia-ESO Consortium} 2013, The Messenger, \href
  {http://adsabs.harvard.edu/abs/2013Msngr.154...47R} {154, 47}

\bibitem[\protect\citeauthoryear{{Revaz}, {Arnaudon}, {Nichols}, {Bonvin}  \&
  {Jablonka}}{{Revaz} et~al.}{2016}]{2016A&A...588A..21R}
{Revaz} Y.,  {Arnaudon} A.,  {Nichols} M.,  {Bonvin} V.,   {Jablonka} P.,
  2016, \mn@doi [\aap] {10.1051/0004-6361/201526438}, \href
  {http://adsabs.harvard.edu/abs/2016A%26A...588A..21R} {588, A21}

\bibitem[\protect\citeauthoryear{{Rosen} \& {Bregman}}{{Rosen} \&
  {Bregman}}{1995}]{1995ApJ...440..634R}
{Rosen} A.,  {Bregman} J.~N.,  1995, \mn@doi [\apj] {10.1086/175303}, \href
  {http://adsabs.harvard.edu/abs/1995ApJ...440..634R} {440, 634}

\bibitem[\protect\citeauthoryear{{Ruiz-Lara}, {Few}, {Gibson}, {P{\'e}rez},
  {Florido}, {Minchev}  \& {S{\'a}nchez-Bl{\'a}zquez}}{{Ruiz-Lara}
  et~al.}{2016}]{2016A&A...586A.112R}
{Ruiz-Lara} T.,  {Few} C.~G.,  {Gibson} B.~K.,  {P{\'e}rez} I.,  {Florido} E.,
  {Minchev} I.,   {S{\'a}nchez-Bl{\'a}zquez} P.,  2016, \mn@doi [\aap]
  {10.1051/0004-6361/201526470}, \href
  {http://adsabs.harvard.edu/abs/2016A%26A...586A.112R} {586, A112}

\bibitem[\protect\citeauthoryear{{Salpeter}}{{Salpeter}}{1955}]{1955ApJ...121..161S}
{Salpeter} E.~E.,  1955, \mn@doi [\apj] {10.1086/145971}, \href
  {http://adsabs.harvard.edu/abs/1955ApJ...121..161S} {121, 161}

\bibitem[\protect\citeauthoryear{{S{\'a}nchez-Bl{\'a}zquez}, {Courty}, {Gibson}
   \& {Brook}}{{S{\'a}nchez-Bl{\'a}zquez} et~al.}{2009}]{2009MNRAS.398..591S}
{S{\'a}nchez-Bl{\'a}zquez} P.,  {Courty} S.,  {Gibson} B.~K.,   {Brook} C.~B.,
  2009, \mn@doi [\mnras] {10.1111/j.1365-2966.2009.15133.x}, \href
  {http://adsabs.harvard.edu/abs/2009MNRAS.398..591S} {398, 591}

\bibitem[\protect\citeauthoryear{{Scannapieco}, {Tissera}, {White}  \&
  {Springel}}{{Scannapieco} et~al.}{2005}]{2005MNRAS.364..552S}
{Scannapieco} C.,  {Tissera} P.~B.,  {White} S.~D.~M.,   {Springel} V.,  2005,
  \mn@doi [\mnras] {10.1111/j.1365-2966.2005.09574.x}, \href
  {http://adsabs.harvard.edu/abs/2005MNRAS.364..552S} {364, 552}

\bibitem[\protect\citeauthoryear{{Scannapieco} et~al.,}{{Scannapieco}
  et~al.}{2012}]{2012MNRAS.423.1726S}
{Scannapieco} C.,  et~al., 2012, \mn@doi [\mnras]
  {10.1111/j.1365-2966.2012.20993.x}, \href
  {http://adsabs.harvard.edu/abs/2012MNRAS.423.1726S} {423, 1726}

\bibitem[\protect\citeauthoryear{{Schmidt}}{{Schmidt}}{1959}]{1959ApJ...129..243S}
{Schmidt} M.,  1959, \mn@doi [\apj] {10.1086/146614}, \href
  {http://adsabs.harvard.edu/abs/1959ApJ...129..243S} {129, 243}

\bibitem[\protect\citeauthoryear{{Serenelli}, {Bergemann}, {Ruchti}  \&
  {Casagrande}}{{Serenelli} et~al.}{2013}]{2013MNRAS.429.3645S}
{Serenelli} A.~M.,  {Bergemann} M.,  {Ruchti} G.,   {Casagrande} L.,  2013,
  \mn@doi [\mnras] {10.1093/mnras/sts648}, \href
  {http://adsabs.harvard.edu/abs/2013MNRAS.429.3645S} {429, 3645}

\bibitem[\protect\citeauthoryear{{Skrutskie} et~al.,}{{Skrutskie}
  et~al.}{2006}]{2006AJ....131.1163S}
{Skrutskie} M.~F.,  et~al., 2006, \mn@doi [\aj] {10.1086/498708}, \href
  {http://adsabs.harvard.edu/abs/2006AJ....131.1163S} {131, 1163}

\bibitem[\protect\citeauthoryear{{Smiljanic} et~al.,}{{Smiljanic}
  et~al.}{2014}]{2014A&A...570A.122S}
{Smiljanic} R.,  et~al., 2014, \mn@doi [\aap] {10.1051/0004-6361/201423937},
  \href {http://adsabs.harvard.edu/abs/2014A%26A...570A.122S} {570, A122}

\bibitem[\protect\citeauthoryear{{Sommer-Larsen} \& {Fynbo}}{{Sommer-Larsen} \&
  {Fynbo}}{2008}]{2008MNRAS.385....3S}
{Sommer-Larsen} J.,  {Fynbo} J.~P.~U.,  2008, \mn@doi [\mnras]
  {10.1111/j.1365-2966.2007.12618.x}, \href
  {http://adsabs.harvard.edu/abs/2008MNRAS.385....3S} {385, 3}

\bibitem[\protect\citeauthoryear{{Springel}}{{Springel}}{2010}]{2010MNRAS.401..791S}
{Springel} V.,  2010, \mn@doi [\mnras] {10.1111/j.1365-2966.2009.15715.x},
  \href {http://adsabs.harvard.edu/abs/2010MNRAS.401..791S} {401, 791}

\bibitem[\protect\citeauthoryear{{Steinmetz} \& {Muller}}{{Steinmetz} \&
  {Muller}}{1995}]{1995MNRAS.276..549S}
{Steinmetz} M.,  {Muller} E.,  1995, \mn@doi [\mnras]
  {10.1093/mnras/276.2.549}, \href
  {http://adsabs.harvard.edu/abs/1995MNRAS.276..549S} {276, 549}

\bibitem[\protect\citeauthoryear{{Stonkut{\.e}} et~al.,}{{Stonkut{\.e}}
  et~al.}{2016}]{2016MNRAS.460.1131S}
{Stonkut{\.e}} E.,  et~al., 2016, \mn@doi [\mnras] {10.1093/mnras/stw1011},
  \href {http://adsabs.harvard.edu/abs/2016MNRAS.460.1131S} {460, 1131}

\bibitem[\protect\citeauthoryear{{Tasker}, {Brunino}, {Mitchell}, {Michielsen},
  {Hopton}, {Pearce}, {Bryan}  \& {Theuns}}{{Tasker}
  et~al.}{2008}]{2008MNRAS.390.1267T}
{Tasker} E.~J.,  {Brunino} R.,  {Mitchell} N.~L.,  {Michielsen} D.,  {Hopton}
  S.,  {Pearce} F.~R.,  {Bryan} G.~L.,   {Theuns} T.,  2008, \mn@doi [\mnras]
  {10.1111/j.1365-2966.2008.13836.x}, \href
  {http://adsabs.harvard.edu/abs/2008MNRAS.390.1267T} {390, 1267}

\bibitem[\protect\citeauthoryear{{Teyssier}}{{Teyssier}}{2002}]{2002A&A...385..337T}
{Teyssier} R.,  2002, \mn@doi [\aap] {10.1051/0004-6361:20011817}, \href
  {http://adsabs.harvard.edu/abs/2002A%26A...385..337T} {385, 337}

\bibitem[\protect\citeauthoryear{{Teyssier}, {Pontzen}, {Dubois}  \&
  {Read}}{{Teyssier} et~al.}{2013}]{2013MNRAS.429.3068T}
{Teyssier} R.,  {Pontzen} A.,  {Dubois} Y.,   {Read} J.~I.,  2013, \mn@doi
  [\mnras] {10.1093/mnras/sts563}, \href
  {http://adsabs.harvard.edu/abs/2013MNRAS.429.3068T} {429, 3068}

\bibitem[\protect\citeauthoryear{{Timmes}, {Woosley}  \& {Weaver}}{{Timmes}
  et~al.}{1995}]{1995ApJS...98..617T}
{Timmes} F.~X.,  {Woosley} S.~E.,   {Weaver} T.~A.,  1995, \mn@doi [\apjs]
  {10.1086/192172}, \href {http://adsabs.harvard.edu/abs/1995ApJS...98..617T}
  {98, 617}

\bibitem[\protect\citeauthoryear{{Tornatore}, {Borgani}, {Matteucci}, {Recchi}
  \& {Tozzi}}{{Tornatore} et~al.}{2004}]{2004MNRAS.349L..19T}
{Tornatore} L.,  {Borgani} S.,  {Matteucci} F.,  {Recchi} S.,   {Tozzi} P.,
  2004, \mn@doi [\mnras] {10.1111/j.1365-2966.2004.07689.x}, \href
  {http://adsabs.harvard.edu/abs/2004MNRAS.349L..19T} {349, L19}

\bibitem[\protect\citeauthoryear{{Turk}, {Smith}, {Oishi}, {Skory}, {Skillman},
  {Abel}  \& {Norman}}{{Turk} et~al.}{2011}]{2011ApJS..192....9T}
{Turk} M.~J.,  {Smith} B.~D.,  {Oishi} J.~S.,  {Skory} S.,  {Skillman} S.~W.,
  {Abel} T.,   {Norman} M.~L.,  2011, \mn@doi [\apjs]
  {10.1088/0067-0049/192/1/9}, \href
  {http://adsabs.harvard.edu/abs/2011ApJS..192....9T} {192, 9}

\bibitem[\protect\citeauthoryear{{Weiss} \& {Schlattl}}{{Weiss} \&
  {Schlattl}}{2008}]{2008Ap&SS.316...99W}
{Weiss} A.,  {Schlattl} H.,  2008, \mn@doi [\apss] {10.1007/s10509-007-9606-5},
  \href {http://adsabs.harvard.edu/abs/2008Ap%26SS.316...99W} {316, 99}

\bibitem[\protect\citeauthoryear{{Wiersma}, {Schaye}  \& {Theuns}}{{Wiersma}
  et~al.}{2011}]{2011MNRAS.415..353W}
{Wiersma} R.~P.~C.,  {Schaye} J.,   {Theuns} T.,  2011, \mn@doi [\mnras]
  {10.1111/j.1365-2966.2011.18709.x}, \href
  {http://adsabs.harvard.edu/abs/2011MNRAS.415..353W} {415, 353}

\bibitem[\protect\citeauthoryear{{Woosley} \& {Weaver}}{{Woosley} \&
  {Weaver}}{1995}]{1995ApJS..101..181W}
{Woosley} S.~E.,  {Weaver} T.~A.,  1995, \mn@doi [\apjs] {10.1086/192237},
  \href {http://adsabs.harvard.edu/abs/1995ApJS..101..181W} {101, 181}

\bibitem[\protect\citeauthoryear{{Yates}, {Henriques}, {Thomas}, {Kauffmann},
  {Johansson}  \& {White}}{{Yates} et~al.}{2013}]{2013MNRAS.435.3500Y}
{Yates} R.~M.,  {Henriques} B.,  {Thomas} P.~A.,  {Kauffmann} G.,  {Johansson}
  J.,   {White} S.~D.~M.,  2013, \mn@doi [\mnras] {10.1093/mnras/stt1542},
  \href {http://adsabs.harvard.edu/abs/2013MNRAS.435.3500Y} {435, 3500}

\bibitem[\protect\citeauthoryear{{van den Hoek} \& {Groenewegen}}{{van den
  Hoek} \& {Groenewegen}}{1997}]{1997A&AS..123..305V}
{van den Hoek} L.~B.,  {Groenewegen} M.~A.~T.,  1997, \mn@doi [\aaps]
  {10.1051/aas:1997162}, \href
  {http://adsabs.harvard.edu/abs/1997A%26AS..123..305V} {123}

\makeatother
\end{thebibliography}
\bibliographystyle{mnras}

\vspace{1cm}
\noindent\small{\textit{$^1$Jeremiah Horrocks Institute, University of Central Lancashire, Preston, PR1~2HE, UK\\
$^{2}$E.A. Milne Centre for Astrophysics, University of Hull, Hull, HU6~7RX, UK\\
$^{3}$Institute for Computational Astrophysics, Dept of Astronomy \& Physics, Saint Mary's University, Halifax, BH3~3C3, Canada\\
$^{4}$Joint Institute for Nuclear Astrophysics - Center for the Evolution of the Elements (JINA-CEE)}}\\
$^{5}$Department of Physics \& Astronomy, University of Exeter, Exeter, EX4~4QL, UK\\
$^{6}$Max-Planck Institute for Astronomy, K{\"o}nigstuhl 17, 69117 Heidelberg, Germany\\
$^{7}$Institute of Space Sciences, Carrer de Can Magrans, Barcelona, E-08193, Spain \\
$^{8}$Institute of Astronomy, University of Cambridge, Madingley Road, Cambridge CB3 0HA, United Kingdom \\
$^{9}$INAF - Osservatorio Astrofisico di Arcetri, Largo E. Fermi 5, 50125, Florence, Italy \\
$^{10}$INAF - Padova Observatory, Vicolo dell'Osservatorio 5, 35122 Padova, Italy \\
$^{11}$Instituto de Astrof\'{i}sica de Andaluc\'{i}a-CSIC, Apdo. 3004, 18080, Granada, Spain \\
$^{12}$Lund Observatory, Department of Astronomy and Theoretical Physics, Box 43, SE-221 00 Lund, Sweden \\
$^{13}$GEPI, Observatoire de Paris, CNRS, Universit\'e Paris Diderot, 5 Place Jules Janssen, 92190 Meudon, France \\
$^{14}$Department of Physics and Astronomy, Uppsala University, Box 516, SE-751 20 Uppsala, Sweden \\
$^{15}$Instituto de F\'isica y Astronomi\'ia, Universidad de Valparai\'iso, Chile \\
$^{16}$European Southern Observatory, Alonso de Cordova 3107 Vitacura, Santiago de Chile, Chile \\
$^{17}$INAF - Osservatorio Astronomico di Bologna, via Ranzani 1, 40127, Bologna, Italy \\
$^{18}$INAF - Osservatorio Astrofisico di Catania, via S. Sofia 78, 95123, Catania, Italy \\
$^{19}$Universit\`e C\^ote d'Azur, OCA, CNRS, Laboratoire Lagrange, CS 34229,F-06304 Nice cedex 4, France \\
$^{20}$Universit\`a di Catania, Dipartimento di Fisica e Astronomia, Sezione Astrofisica, Via S. Sofia 78, I-95123 Catania, Italy \\
$^{21}$Astrophysics Research Institute, Liverpool John Moores University, 146 Brownlow Hill, Liverpool L3 5RF, United Kingdom \\
$^{22}$Universit\`e C$\hat{o}$te d'Azur, Observatoire de la C$\hat{o}$te d'Azur, CNRS, Laboratoire Lagrange, Bd de l'Observatoire, CS 34229,\\06304 Nice Cedex 4, France \\
$^{23}$Departamento de Ciencias Fisicas, Universidad Andres Bello, Republica 220, Santiago, Chile \\
$^{24}$INAF - Osservatorio Astronomico di Palermo, Piazza del Parlamento 1, 90134, Palermo, Italy \\
$^{25}$Instituto de Astrof\'isica e Ci\^encias do Espa\c{c}o, Universidade do Porto, CAUP, Rua das Estrelas, 4150-762 Porto, Portugal \\
$^{26}$Institute of Theoretical Physics and Astronomy, Vilnius University, Sauletekio av. 3, 10222 Vilnius, Lithuania \\

\label{lastpage}
\end{document}